\documentclass[twoside,english,sort&compress]{iopart}
\usepackage{mathptmx}

\usepackage[T1]{fontenc}
\usepackage[latin9]{inputenc}
\usepackage{geometry}
\usepackage{xcolor}
\usepackage{units}
\usepackage{amsbsy}
\usepackage{amstext}
\usepackage{graphicx}
\usepackage[numbers]{natbib}

\makeatletter

\providecommand{\tabularnewline}{\\}

\usepackage{iopams}
\usepackage{setstack}

\newcommand{\eqref}[1]{(\ref{#1})}
\makeatother

\usepackage{babel}
\begin{document}

\title[Generalized Fisher-Lee relation]{Equivalence of wave function matching and Green's functions methods
for quantum transport: generalized Fisher-Lee relation}

\author{{\Large{}Hocine Boumrar$^{1}$, Mahdi Hamidi$^{1}$, Hand Zenia$^{1}$
and Samir Lounis$^{2}$} }

\address{{\Large{}$^{1}$}{\large{}Laboratoire de Physique et Chimie Quantique
(LPCQ), Université Mouloud Mammeri, 15000 Tizi-ouzou, Algeria}\\
{\Large{}$^{2}$}{\large{}Peter Grünberg Institut and Institute for
Advanced Simulation, Forschungszentrum Jülich \& JARA, 52425 Jülich,
Germany}}

\ead{{\large{}h.zenia@gmail.com}}
\begin{abstract}
We present a proof of an exact equivalence of the two approaches that
are most used in computing conductance in quantum electron and phonon
transport: the wave function matching and Green's functions methods.
We can obtain all the quantities defined in one method starting from
those obtained in the other. This completes and illuminates the work
started Ando{[}Ando T 1991 \textit{Phys. Rev.} B \textbf{44} 8017.{]}
and continued later by Khomyakov \textit{et al}.{[}Khomyakov P A ,
Brocks G, Karpan V, Zwierzycki M and Kelly P J 2005 \textit{Phys.
Rev.} B \textbf{72} 035450.{]}.\textcolor{brown}{{} }The aim is to allow
for solving the transport problem with whichever approach fits most
the system at hand. One major corollary of the proven equivalence
is our derivation of a generalized Fisher-Lee formula for resolving
the transmission function into individual phonon mode contributions.
As an illustration, we applied our method to a simple model to highlight
its accuracy and simplicity.
\end{abstract}


\submitto{\JPCM }
\maketitle

\section{Introduction}

Electron and phonon transport in mesoscopic systems has gained much
interest in the last three decades, due mainly to advances in experimental
and computational techniques, as well as the continued improvements
in computing resources\citep{datta02,diventra08}. Indeed, experimental
techniques are nowadays so precise, it has become feasible to wire
a nanometric-sized molecule to two metallic leads and measure its
transport properties\citep{evers19}. The technological applications
are also varied as a result of the trend to miniaturization, but also
the push to exploit the novel effects observed only at the atomic
scale. Heat dissipation, Josephson junctions, thermoelectricity, spintronics
are among the most prominent applications that rely on such novel
phenomena \citep{Bell08,Dubi2011,chahil2014,freericks06}. In general,
the setup consists of a central part, the device, considered small,
connected to much larger leads. The leads act as heat and particle
reservoirs each at a constant temperature an chemical potential. When
the leads are kept at slightly different temperatures or chemical
potentials, a heat or particle current ensues that passes through
the central region. The first attempts at a theoretical understanding
of the phenomenon were made in the second half of the last century.
We note that when interactions are involved, the heat current in the
electronic case is harder to describe. However, in this work we are
concerned with the heat transport via harmonic phonons, whose treatment
is also similar to the electronic charge transport. Using a phenomenological
approach, Landauer\citep{landauer57,landauer70} expressed the conductance
in a two-terminal setup in terms of a transmission probability. There
are mainly two methods that are used to compute the latter. The first
relies on using wave functions in the electronic case or displacements
in the phononic case, and is sometimes known as the wave function
matching (WFM) method\citep{ando91,khomyakov04,xia06,farmanbar16,chen19}.
The second uses the Green's function as the main object, and is known
as the atomistic Green's functions method (AGF)\citep{datta02,Mingo2003,Wang2008,Zhang2007,Li2012,Gu2015,Tian2012}.

The AGF was first applied to transport by Caroli \textit{et al.}\citep{caroli71}
using a Hamiltonian with a basis of localized wave functions. Their
aim was to circumvent the inherently difficult problem of matching
correctly the wave functions at the interfaces. Moreover, Green's
functions are the most convenient tool to include correlations in
a more realistic treatment of the problem when these are important\citep{freericks06}.
The method is also ``unique'' in that studying systems far from
thermal equilibrium relies exclusively on the use of Green's functions
as introduced by the Keldysh formalism \citep{diventra08,Keldysh1965,Zhu2016}.
An alternative to the Green's function method, consists of studying
the scattering problem using wave functions. This requires that appropriate
boundary conditions be imposed at the interfaces between the scattering
central region and the leads. Ando\citep{ando91} applied the wave
function matching method to a tight-binding Hamiltonian, obtaining
directly the transmission as well as reflexion probability amplitudes.
Ando's approach was criticized by Krsti\'c \textit{et al.}\citep{krstic02}
for not yielding the correct expression of the conductance, due mainly,
as their argument goes, to an inconsistent treatment of the evanescent
modes. It is indeed the case that these modes do not participate in
the transport, but their inclusion is necessary for a correct matching
of the wave functions at the interfaces. Khomyakov \textit{et al.}\citep{khomyakov05}
later attempted to demonstrate the equivalence between the wave function
matching and Green's functions methods and show that Ando's approach
was correct. Their attempt was successful at deriving the Green's
functions from quantities obtained using the wave functions method.
However, finding the complete equivalence remained an open question.
Beyond the mere challenge of showing that an equivalence exists, the
fallout from such a demonstration is substantial in that one can readily
check whether a calculation carried out using either method is corroborated
by the other. This is indeed a much needed way to reduce errors, given
that the problems tackled are way beyond reach of analytical tools\textit{\textcolor{blue}{.
}}\textcolor{black}{Moreover, for more realistic calculations one
has to appeal to first-principles methods based mostly on the density
functional theory(DFT). Most computer codes that implement the DFT
make use of the wave function formalism, whereby the appropriate Khon-Sham
equations are solved by expanding the electronic wave function on
a basis of either localized or extended wave functions\citep{Giannozzi2009}.
A few other codes rely on the Green's function formalism\citep{Ebert2011}.
It is, therefore, convenient to implement transport calculations on
top of the existing codes using either the WFM or AGF method, depending
on the nature of the code, wave-function-based or Green's-functions-based,
and knowing that the results will be identical in either case. }

Besides the total transmission function, one major aspect of the whole
framework is obtaining mode-resolved transmission probabilities, i.e.,
individual phonon contributions to the total transmission function
\citep{Huang2011,Sadasivan2017,Ong2015,Latour2017,Ong2018,Klochner2018}.
These are important for several reasons. For instance, such detailed
information is required as input to other methods used at longer length
scales, such as, non-gray Boltzmann equation \citep{Huang2011}. One
can also consider the interaction between phonons and light, where
only the optical branches intervene, hence the need to disentangle
their contribution from the rest of the phonon spectrum. The first
attempt at deriving these mode-resolved transmissions was contained
in the Fisher-Lee relation \citep{Fisher1981}. Several attempts have
been made later on at obtaining such information, but most were misled
by the invalid assumption that for a fixed frequency, the eigenvectors
are orthogonal \citep{datta02,Ferry2001}. We derive, in the current
work, the correct and physically sound generalized Fisher-Lee relation
for the mode-resolved transmission probabilities. Aside from the ease
of implementation, the final expression which we arrived at has a
clear physical interpretation. Previous derivations were obtained
through rather involved mathematical machinery\citep{ando91,khomyakov05,Sanvito1999,Wimmer2009,zhang2017},
and the final expressions, though amenable to our result, still require
further manipulations. 

In this work we provide a rigorous demonstration of the equivalence
of the WFM and the AGF methods by showing that all the quantities
calculated within one approach can be readily obtained from the other.
These include, for instance, group velocities, transmission probabilities,
and the individual phonon contributions to these probabilities. In
order to illustrate the concordance, we compare various quantities
computed within the two frameworks applied to a simple model. Moreover,
we also argue that in order to describe correctly the problem at hand
the usual region defined as the central part must include one more
atomic plane from each lead.

This paper is organized as follows: In the next section our WFM formalism
is introduced and the transmission function is derived within the
formalism. In section 3 we provide the equivalence between the WFM
and the AGF methods, and derive a generalized Fisher-Lee formula for
the mode-resolved transmission amplitudes. In section 4 we apply our
method to a simple model and make comparisons with earlier approaches.
We finally summarize our main results and discussions in the conclusion. 

\section{Wave function matching approach}

\begin{figure}
\begin{centering}
\includegraphics[scale=0.5]{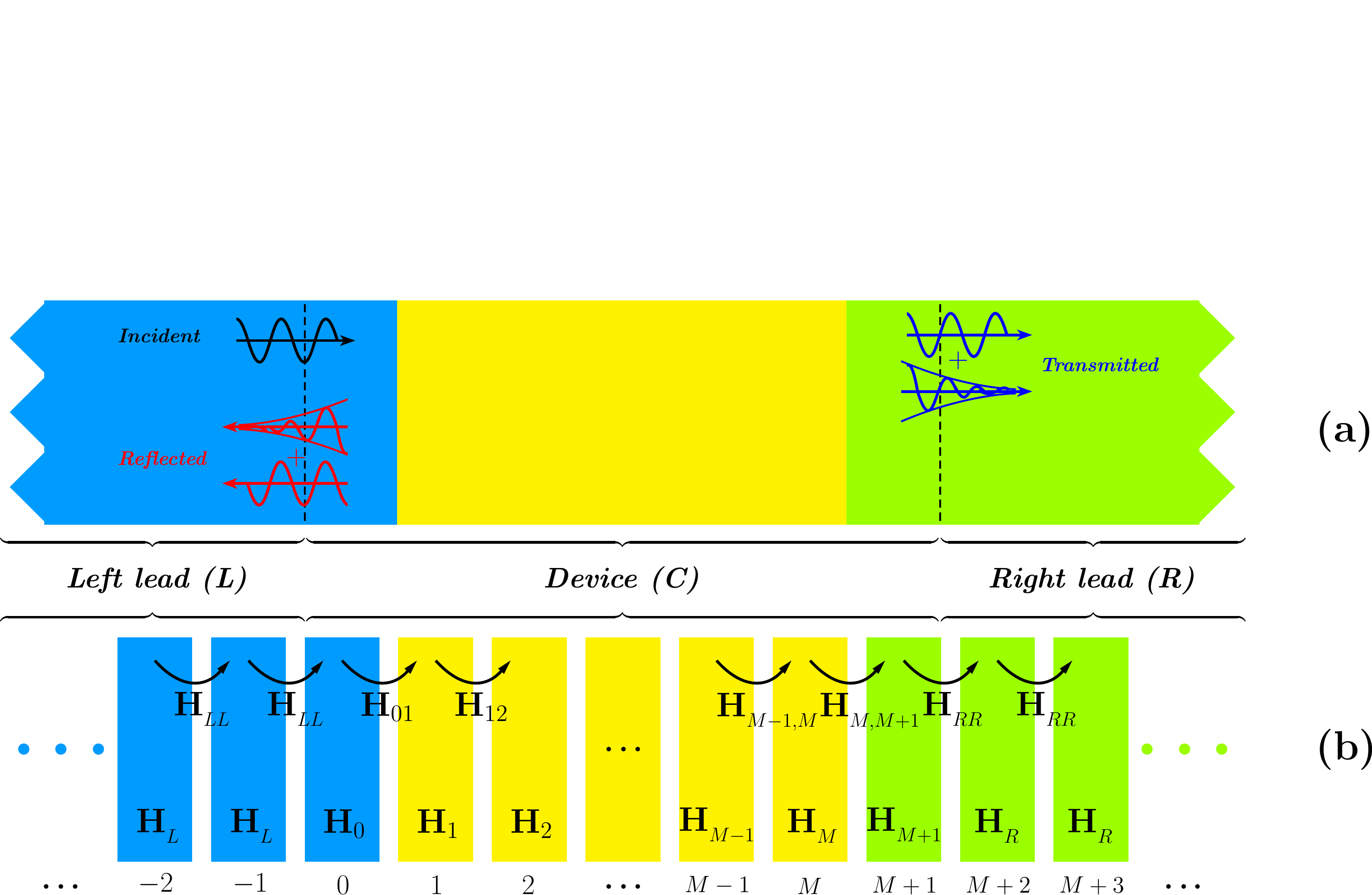}
\par\end{centering}
\caption{(a) Schematic representation and (b) Hamiltonian matrix divided into
slices for the bi-dimensional nano-scale system studied within the
WFM approach.\label{fig_device}}
\end{figure}

Our main objective is to provide a full correspondence of the two
approaches mentioned earlier. We think that the Green's functions
formalism is the more robust, and as such we posit that all the quantities
defined in this formalism must be expressible in terms of variables
defined in the wave function matching approach. Indeed, the AGF method
suffers no ambiguity in defining and computing the various quantities
involved in the transport problem. The WFM approach, on the other
hand, has been subject to debate as to its correct implementation
in the present situation. One chief contention regards the treatment
of the decaying modes that result from the presence of the scattering
region. In the AGF approach the two main quantities are the retarded
and advanced Green's functions. The first describes outgoing waves
from the device, while the second is related to incoming waves. From
the mathematical standpoint, an in order to obtain these two quantities
from the WFM formalism, we must keep only the evanescent modes that
decay away from the device. There is also a compelling physical argument
in favor of this choice: the evanescent modes always originate from
the device, and can only decay away from it. Besides, if we were to
reproduce the asymptotic behavior of the transmission probability,
we must not include incoming evanescent modes since they would never
reach the device if we inject them far away from it, as is clearly
illustrated in Fig. \ref{fig_device}.

We begin by reminding the formalism used in studying the scattering
problem using matching of the wave functions at the interfaces of
a bi-dimensional nano-scale system composed of semi-infinite right
and left leads separated with a central device (see Figure 1-a). The
leads are considered to be ideal with a periodic potential. The incoming
propagating state from the left lead and normal to the interface (say
in $x$ direction) is scattered in the device and results in reflected
and transmitted propagating as well as evanescent states. In the transverse
direction (normal to $x$) the periodicity is assumed to hold. Consequently,
the real space components in the transverse direction can be replaced
by the normal component of the momentum ($\mathbf{k}_{\perp}$). As
is usual we first derive quantities proper to the ideal left and right
leads, to be used later to express the transmission probabilities.

The dynamical matrix of the whole system is given by the following
infinite tridiagonal matrix:

\begin{equation}
\mathbf{H}=\left[\begin{array}{ccccccc}
\ddots & \ddots & 0 & 0 & 0 & 0 & 0\\
\ddots & \mathbf{H}_{L} & \mathbf{H}_{LL} & 0 & 0 & 0 & 0\\
0 & \mathbf{H}_{LL}^{\dagger} & \mathbf{H}_{L} & \mathbf{H}_{LL} & 0 & 0 & 0\\
0 & 0 & \mathbf{H}_{LL}^{\dagger} & \mathbf{H}_{C} & \mathbf{H}_{RR} & 0 & 0\\
0 & 0 & 0 & \mathbf{H}_{RR}^{\dagger} & \mathbf{H}_{R} & \mathbf{H}_{RR} & 0\\
0 & 0 & 0 & 0 & \mathbf{H}_{RR}^{\dagger} & \mathbf{H}_{R} & \ddots\\
0 & 0 & 0 & 0 & 0 & \ddots & \ddots
\end{array}\right],\label{006}
\end{equation}
where the indices $L$ and $R$ stand for the left and right leads,
respectively, as shown in Fig. \ref{fig_device}(b). The part $\mathbf{H}_{C}$
represents the scattering region for which the Hamiltonian matrix
is given as
\begin{equation}
\mathbf{H}_{C}=\left[\begin{array}{ccccccc}
\mathbf{H}_{0} & \mathbf{H}_{01} & 0 & 0 & 0 & 0 & 0\\
\mathbf{H}_{01}^{\dagger} & \mathbf{H}_{1} & \mathbf{H}_{12} & 0 & 0 & 0 & 0\\
0 & \mathbf{H}_{12}^{\dagger} & \mathbf{H}_{2} & \ddots & 0 & 0 & 0\\
0 & 0 & \ddots & \ddots & \ddots & 0 & 0\\
0 & 0 & 0 & \ddots & \mathbf{H}_{M-1} & \mathbf{H}_{M-1,M} & 0\\
0 & 0 & 0 & 0 & \mathbf{H}_{M-1,M}^{\dagger} & \mathbf{H}_{M} & \mathbf{H}_{M,M+1}\\
0 & 0 & 0 & 0 & 0 & \mathbf{H}_{M,M+1}^{\dagger} & \mathbf{H}_{M+1}
\end{array}\right].\label{007}
\end{equation}
Here the slides $0$ and $M+1$ are included in the scattering region,
although their respective Hamiltonians are seemingly similar to the
left and right leads. Indeed, when either the force constants or the
masses of the three regions are different, the Hamiltonians $\mathbf{H}_{0}$
and $\mathbf{H}_{M+1}$ are different than $\mathbf{H}_{L}$ and $\mathbf{H}_{R}$,
respectively. Because of the assumption that any given slice is connected
only to its immediate neighboring slices, the dynamical matrix takes
on a tri-diagonal form. As such one atomic slice can be made up of
one or more atomic planes, depending on whether interactions are first-neighbor
only or are of a longer range (see Fig. \ref{fig_device}(b)). The
left and right leads are semi-infinite and extend all the way to $-\infty$
and to $+\infty$, respectively.

\subsection{Ideal leads}

From now on we restrict ourselves to studying the phonon transport,
although all the ensuing discussions can be easily carried on to the
electronic charge transport. In order to study transport of phonons
from one side of the device to the other, we need to first study the
motion of phonons in the ideal leads, that is in each lead as if it
were infinite in both directions. The eigenmodes of these ideal leads
will then be used to study how a phonon sent from the left lead scatters
off the central part and results in transmitted and reflected phonons.
We begin, therefor, by obtaining the eigenmodes of the ideal leads.

As shown in Figure1-b, the studied system consists of three parts
of slices normal to the direction of propagation and indexed from$-\infty$
to $-1$ for the left lead, from $0$ to $M+1$ for the scattering
region, and from $M+2$ to $+\infty$ for the right lead. Here we
note that one slice from each lead is included as part of the scattering
region. The slices are chosen such that only the interactions between
neighboring slices are considered.

The equations of motion for the atoms in each ideal lead are given
by
\begin{equation}
\omega^{2}\mathbf{u}_{S,i}=\mathbf{H}_{SS}^{\dagger}\,\mathbf{u}_{S,i-1}+\mathbf{H}_{S}\,\mathbf{u}_{S,i}+\mathbf{H}_{SS}\,\mathbf{u}_{S,i+1},\label{eq:008}
\end{equation}
where $\mathbf{H}_{S}$ and $\mathbf{H}_{SS}$ ($S\equiv R$ for the
right lead and $S\equiv L$ for the left lead) represent the on-slice
and the inter-slice parts of the Hamiltonian, respectively. In this
case $i$ runs from $-\infty$ to $+\infty$. It is important to bear
in mind that all the Hamiltonian and force constant matrices depend
explicitly on $k_{\perp}$, i.e the momentum component which is transverse
to the propagation direction $x$. We choose not to show this dependence
for the sake of uncluttering the formulae.

They are $N\times N$ matrices, where $N$ is the number of ``vibrational
degrees of freedom'' per slice. In the electronic problem, $N$ is
the number of orbitals per slice used as a basis to represent the
electronic wave functions. The component of a vibrational mode in
slice $i$ is represented by a vector $\mathbf{u}_{S,n}$ of size
$N$. The equations of motion can be cast in matrix form as

\begin{equation}
\omega^{2}\left(\begin{array}{c}
\vdots\\
\mathbf{u}_{S,i-1}\\
\mathbf{u}_{S,i}\\
\mathbf{u}_{S,i+1}\\
\vdots
\end{array}\right)=\left[\begin{array}{ccccc}
\ddots & \ddots & 0 & 0 & 0\\
\ddots & \mathbf{H}_{S} & \mathbf{H}_{SS} & 0 & 0\\
0 & \mathbf{H}_{SS}^{\dagger} & \mathbf{H}_{S} & \mathbf{H}_{SS} & 0\\
0 & 0 & \mathbf{H}_{SS}^{\dagger} & \mathbf{H}_{S} & \ddots\\
0 & 0 & 0 & \ddots & \ddots
\end{array}\right]\left(\begin{array}{c}
\vdots\\
\mathbf{u}_{S,i-1}\\
\mathbf{u}_{S,i}\\
\mathbf{u}_{S,i+1}\\
\vdots
\end{array}\right),\label{eq:009}
\end{equation}
For the ideal right lead, where the same unit cell or slice is repeated
indefinitely in the $x$ direction of space, we make use of Bloch's
theorem and write ${\displaystyle {\displaystyle \mathbf{u}_{R,i}=e^{iq_{x}\,a_{R}}\,\mathbf{u}_{R,i-1}}}$,
where $q_{x}$ and $a_{R}$ are the wave vector and the size of the
unit cell in the $x$ direction, respectively. It is important to
stress, at this point, that $q_{x}$ can be either a purely real or
a complex, having both real and imaginary parts, number. In the latter
case the use of the expression ``Bloch's theorem'' is a misnomer
given that the corresponding modes are not propagating. Defining ${\displaystyle \lambda_{R}=e^{iq_{x}\,a_{R}}}$,
we have $\mathbf{u}_{R,i}=\lambda_{R}\,\mathbf{u}_{R,i-1}$. We can
now rewrite the equations of motion \eqref{eq:008} as 
\begin{equation}
\omega^{2}\mathbf{u}_{R,i}=\mathbf{H}_{RR}^{\dagger}\,\lambda_{R}^{-1}\,\mathbf{u}_{R,i}+\mathbf{H}_{R}\,\mathbf{u}_{R,i}+\mathbf{H}_{RR}\,\lambda_{R}\,\mathbf{u}_{R,i}.\label{eq:010}
\end{equation}
That is
\begin{equation}
\left(\omega^{2}-\mathbf{H}_{R}-\lambda_{R}^{-1}\,\mathbf{H}_{RR}^{\dagger}-\lambda_{R}\,\mathbf{H}_{RR}\right)\mathbf{u}_{R,i}=0.\label{eq:011}
\end{equation}
This is a quadratic eigenvalue problem which can be recast as a generalized eigenvalue problem by
defining $\lambda_{R}^{-1}\,\mathbf{u}_{R}$ as a new ``independent''
variable. The new generalized eigenvalue problem then reads
\begin{equation}
\left[\begin{array}{cc}
\omega^{2}\mathbf{I}-\mathbf{H}_{R} & -\mathbf{H}_{RR}^{\dagger}\\
\mathbf{I} & 0
\end{array}\right]\left(\begin{array}{c}
\mathbf{u}_{R}\\
\lambda_{R}^{-1}\,\mathbf{u}_{R}
\end{array}\right)=\lambda_{R}\left[\begin{array}{cc}
\mathbf{H}_{RR} & 0\\
0 & \mathbf{I}
\end{array}\right]\left(\begin{array}{c}
\mathbf{u}_{R}\\
\lambda_{R}^{-1}\,\mathbf{u}_{R}
\end{array}\right).\label{eq:012}
\end{equation}
Similarly, for the ideal periodic left lead, we have chosen to write,
using Bloch's theorem, $\mathbf{u}_{L,i-1}=\tilde{\lambda}_{L}\,\mathbf{u}_{L,i}$
, where ${\displaystyle \tilde{\lambda}_{L}=e^{-i\,q_{x}\,a_{L}}}$
and $a_{L}$ is the size of the left unit cell. Let us note that the
choice of ${\displaystyle \tilde{\lambda}_{L}}$ as the Bloch factor
being different from that of the right lead is only motivated by the
aim to obtain results that are homogeneous with those obtained through
the AGF formalism, as it can be seen later. We can now rewrite Eq.
\eqref{eq:008} as 
\begin{equation}
\omega^{2}\mathbf{u}_{L,n}=\mathbf{H}_{LL}^{\dagger}\,\tilde{\lambda}_{L}\,\mathbf{u}_{L,i}+\mathbf{H}_{L}\,\mathbf{u}_{L,i}+\mathbf{H}_{LL}\,\tilde{\lambda}_{L}^{-1}\,\mathbf{u}_{L,i},\label{eq:013}
\end{equation}
and then
\begin{equation}
\left(\omega^{2}-\mathbf{H}_{L}-\tilde{\lambda}_{L}\,\mathbf{H}_{LL}^{\dagger}-\tilde{\lambda}_{L}^{-1}\,\mathbf{H}_{LL}\right)\mathbf{u}_{L,n}=0.\label{eq:014}
\end{equation}
By defining the new ``independent'' variable $\tilde{\lambda}_{L}^{-1}\,\mathbf{u}_{L}$,
the quadratic eigenvalue problem becomes a generalized eigenvalue problem and reads
\begin{equation}
\left[\begin{array}{cc}
\omega^{2}\mathbf{I}-\mathbf{H}_{L} & -\mathbf{H}_{LL}\\
\mathbf{I} & 0
\end{array}\right]\left(\begin{array}{c}
\mathbf{u}_{L}\\
\tilde{\lambda}_{L}^{-1}\,\mathbf{u}_{L}
\end{array}\right)=\tilde{\lambda}_{L}\left[\begin{array}{cc}
\mathbf{H}_{LL}^{\dagger} & 0\\
0 & \mathbf{I}
\end{array}\right]\left(\begin{array}{c}
\mathbf{u}_{L}\\
\tilde{\lambda}_{L}^{-1}\,\mathbf{u}_{L}
\end{array}\right).\label{eq:015}
\end{equation}

From \eqref{eq:012} and \eqref{eq:015}, for each value of $\omega$,
we obtain $2N$ eigenvalues and eigenvectors for each ideal lead.
The eigenmodes are split into two sets of size $N$ each:
\begin{enumerate}
\item Right-moving propagating ($\left|\lambda_{R,n}\right|=1\,\mathrm{or}\,\left|\tilde{\lambda}_{L,n}\right|=1\;\textrm{with}\;v_{R/L,n}>0$)
and evanescent ($\left|\lambda_{R,n}\right|<1\,\mathrm{or}\,\left|\tilde{\lambda}_{L,n}\right|>1$)
modes, where $v_{R/L,n}$ is the group velocity of the $n^{\mathrm{th}}$ mode along the $x$-direction. 
\item Left-moving propagating ($\left|\lambda_{R,n}\right|=1\,\mathrm{or}\,\left|\tilde{\lambda}_{L,n}\right|=1\;\textrm{with}\;v_{R/L,n}<0$) and evanescent ($\left|\lambda_{R,n}\right| >1 \,\mathrm{or}\,\left|\tilde{\lambda}_{L,n}\right|<1$)
modes.
\end{enumerate}
In view of attaching later the leads to the scattering region and
having in sight the equivalence with the AGF formalism, we depart
from the usual treatments \citep{ando91,khomyakov04,xia06,Wang2008}\textcolor{blue}{{}
}and discard the evanescent modes that decay towards the central region.
As mentioned above, the physical justification resides in that evanescent
modes are localized around the ``defect'' and always decay away
from it. It does not make sense to include evanescent modes moving
towards the ``defect''. We also know for a fact that this is indeed
the case when using the Green's functions formalism, and by this clever
trick, we will show how to move seamlessly between the two formalisms. And in keeping with the terminology used in the Green's function formalism, we define quantities built out of these selected evanescent modes and propagating modes moving either towards the defect or away from it. We call the latter ``retarded'' and the first ``advanced'', and in order to differentiate them we append a superscript ``a'' to the advanced quantities. We note that in what follows the advanced quantities for the right lead are defined only for the sake of completeness. So long as the incoming phonon is injected from the left lead, the boundary conditions are imposed such that there are only transmitted waves in the right lead. These waves are represented by the retarded quantities. As such the advanced quantities are not needed in the right lead. However, if one were to inject phonons from the right lead, then the advanced quantities in the  left lead would be the ones to be unused. We begin by reminding the definition of the group velocity:
\begin{equation}
v_{S}=\frac{\partial\omega}{\partial q_{x}}=\frac{1}{2\omega}\frac{\partial\omega^{2}}{\partial q_{x}}.
\end{equation}
From the equations of motion in the leads and using the Hellmann-Feynman
theorem \citep{tannoudji}, we obtain the following expressions
for the various group velocities defined in the right and left leads:
\begin{eqnarray}
v_{g,R} & = & \frac{\mathrm{i}\,a_{R}}{2\omega}\left[\mathbf{u}_{R}^{\dagger}\,\mathbf{H}_{RR}\,\mathbf{u}_{R}\,\lambda_{R}-\mathbf{u}_{R}^{\dagger}\,\mathbf{H}_{RR}^{\dagger}\,\mathbf{u}_{R}\,\lambda_{R}^{\dagger}\right],\\
v_{g,R}^{a} & = & \frac{\mathrm{i}\,a_{L}}{2\omega}\left[\mathbf{u}_{R}^{a\dagger}\,\mathbf{H}_{RR}\,\mathbf{u}_{R}^{a}\,\lambda_{R}^{a}-\mathbf{u}_{R}^{a\dagger}\,\mathbf{H}_{RR}^{\dagger}\,\mathbf{u}_{R}^{a}\,\lambda_{R}^{a\dagger}\right],\\
v_{g,L} & = & -\frac{\mathrm{i}\,a_{L}}{2\omega}\left[\mathbf{u}_{L}^{\dagger}\,\mathbf{H}_{LL}^{\dagger}\,\mathbf{u}_{L}\,\tilde{\lambda}_{L}-\mathbf{u}_{L}^{\dagger}\,\mathbf{H}_{LL}\,\mathbf{u}_{L}\,\tilde{\lambda}_{L}^{\dagger}\right],\\
v_{g,L}^{a} & = & -\frac{\mathrm{i}\,a_{L}}{2\omega}\left[\mathbf{u}_{L}^{a\dagger}\,\mathbf{H}_{LL}^{\dagger}\,\mathbf{u}_{L}^{a}\,\tilde{\lambda}_{L}^{a}-\mathbf{u}_{L}^{a\dagger}\,\mathbf{H}_{LL}\,\mathbf{u}_{L}^{a}\,\tilde{\lambda}_{L}^{a\dagger}\right],
\end{eqnarray}
where $a_{L}$\textcolor{black}{{} and $a_{R}$ are the unit cell size
in the left and right leads, respectively. We also define new quantities
which are constructed out of the group velocities, with the coefficients
${\displaystyle \frac{a_{L/R}}{2\omega}}$ omitted for convenience
when we will be comparing the WFM with the AGF formalisms:}

\textcolor{black}{
\begin{equation}
\mathbf{V}_{R}=\mathrm{i}\left[\mathbf{C}_{R}^{\dagger}\,\mathbf{H}_{RR}\,\mathbf{C}_{R}\,\mathbf{\boldsymbol{\Lambda}}_{R}-\mathbf{\boldsymbol{\Lambda}}_{R}^{\dagger}\,\mathbf{C}_{R}^{\dagger}\,\mathbf{H}_{RR}^{\dagger}\,\mathbf{C}_{R}\right],\label{v_r_matrix}
\end{equation}
}

\textcolor{black}{
\begin{equation}
\mathbf{V}_{R}^{a}=\mathrm{i}\left[\mathbf{C}_{R}^{a\dagger}\,\mathbf{H}_{RR}\,\mathbf{C}_{R}^{a}\,\mathbf{\boldsymbol{\Lambda}}_{R}^{a}-\mathbf{\boldsymbol{\Lambda}}_{R}^{a\dagger}\,\mathbf{C}_{R}^{a\dagger}\,\mathbf{H}_{RR}^{\dagger}\,\mathbf{C}_{R}^{a}\right],
\end{equation}
\begin{equation}
\mathbf{V}_{L}=-\mathrm{i}\left[\mathbf{C}_{L}^{\dagger}\,\mathbf{H}_{LL}^{\dagger}\,\mathbf{C}_{L}\,\mathbf{\boldsymbol{\Lambda}}_{L}-\mathbf{\boldsymbol{\Lambda}}_{L}^{\dagger}\,\mathbf{C}_{L}^{\dagger}\,\mathbf{H}_{LL}\,\mathbf{C}_{L}\right],
\end{equation}
}

\textcolor{black}{
\begin{equation}
\mathbf{V}_{L}^{a}=-\mathrm{i}\left[\mathbf{C}_{L}^{a\dagger}\,\mathbf{H}_{LL}^{\dagger}\,\mathbf{C}_{L}^{a}\,\mathbf{\boldsymbol{\Lambda}}_{L}^{a}-\mathbf{\boldsymbol{\Lambda}}_{L}^{a\dagger}\,\mathbf{C}_{L}^{a\dagger}\,\mathbf{H}_{LL}\,\mathbf{C}_{L}^{a}\right].\label{vla_matrix}
\end{equation}
These matrices are diagonal, with the diagonal elements being the
group velocities for the propagating modes, and vanishing for the
decaying modes.}

We now introduce the matrices built from the eigenvectors and eigenmodes
of the ideal leads that will be used in the WFM formalism to obtain
the vibration modes in the central region. For the right lead, we
define

\begin{equation}
\mathbf{C}_{R}=\left(\mathbf{u}_{R,1},\mathbf{u}_{R,2},\cdots,\mathbf{u}_{R,n}\right),
\end{equation}

\begin{equation}
\mathbf{C^{\mathrm{\mathit{a}}}}_{R}=\left(\mathbf{u}_{R,1}^{a},\mathbf{u}_{R,2}^{a},\cdots,\mathbf{u}_{R,n}^{a}\right),
\end{equation}

\begin{equation}
\boldsymbol{\Lambda}_{R}=\mathrm{diag}\left(\lambda_{R,1},\lambda_{R,2},\cdots,\lambda_{R,n}\right),
\end{equation}
and
\begin{equation}
\boldsymbol{\Lambda}_{R}^{a}=\mathrm{diag}\left(\lambda_{R,1}^{a},\lambda_{R,2}^{a},\cdots,\lambda_{R,n}^{a}\right),
\end{equation}
and we define similar quantities for the left lead with $\lambda_{R}$ replaced by $\tilde{\lambda}_{L}$, and $\mathbf{u}_{R}$ by $\mathbf{u}_{L}$.

\subsection{The scattering region}

We now proceed to writing the effective Hamiltonian of the scattering
region, replacing the effects of the ideal leads by ``self-energy''
terms to be added to the original Hamiltonian.\textcolor{black}{{} Writing
the equation of motion for slice $0$, 
\[
\mathbf{-H}_{LL}^{\dagger}\mathbf{\mathbf{u}_{\mathrm{-1}}\mathbf{+}\mathrm{(}\omega^{2}\mathbf{I}-H}_{0})\mathbf{\mathbf{u}_{\mathrm{0}}-H}_{01}\mathbf{u}_{1}=0,
\]
and moving the first term to the right-hand side, we have
\[
\mathbf{\mathrm{(}\omega^{2}\mathbf{I}-H}_{0})\mathbf{\mathbf{u}_{\mathrm{0}}-H}_{01}\mathbf{u}_{1}=\mathbf{H}_{LL}^{\dagger}\mathbf{u}_{-1}.
\]
In a similar fashion, we find for slice $M+1$,
\[
\mathbf{\mathrm{(}\omega^{2}\mathbf{I}-H}_{M+1})\mathbf{\mathbf{u}_{\mathrm{M+1}}-H}_{M,M+1}\mathbf{u}_{M}=\mathbf{H}_{RR}\mathbf{u}_{M+2}.
\]}\noindent The  equations of motion for the central part are then given in matrix
form as
\begin{equation}
\left[\omega^2\mathbf{I}-\mathbf{H}_C\right]\left[\begin{array}{c}
\mathbf{u}_{0}\\
\mathbf{u}_{1}\\
\\
\vdots\\
\\
\mathbf{u}_{M}\\
\mathbf{u}_{M+1}
\end{array}\right]=\left[\begin{array}{c}
\mathbf{\mathbf{H}_{LL}^{\dagger}u}_{-1}\\
0\\
0\\
\vdots\\
0\\
0\\
\mathbf{\mathbf{H}_{RR}u}_{M+2}
\end{array}\right],\label{central_eom}
\end{equation}
with the full matrix form of $\mathbf{H}_C$ given in Eq. \ref{007}.

\noindent This means that if we know or impose the values of $\mathbf{u}_{-1}$
and $\mathbf{u}_{M+2},$ we can invert this matrix equation to obtain
all the displacement $\mathbf{u}_{i},$ with $i=0,1,...,M+1$. 

To accomplish this, we make use of Bloch's theorem which holds in
the leads. We write \begin{equation}     
    \left\{\begin{array}{lr}    
              \mathbf{u}_{-1}=&{\displaystyle \tilde{\lambda}_{L}\mathbf{u}_{0}}\quad\textrm{with}\quad
                           \tilde{\lambda}=e^{-i\,q_{x}a_{L}}, \\    
              \mathbf{u}_{M+2}=&{\displaystyle \lambda_{R}\,\mathbf{u}_{M+1}}\quad\textrm{with}\quad
                           \lambda_{R}      =e^{i\,q_{x}a_{R}}              
           \end{array}\right.,  
\end{equation}Imposing an incoming mode from the left lead, with an amplitude $\mathbf{a}_{in}$
, the displacement $\mathbf{u}_{0}$ will be a linear combination
of the incoming and reflected waves. In the right lead we impose that
only the transmitted wave exist. We have then \begin{equation}     
    \left\{\begin{array}{lr}         
              \mathbf{u}_{0} =&\mathbf{C}_{L}^{a}\,\mathbf{a}_{in}+\mathbf{C}_{L}\,\mathbf{a}_{ref},\\
              \mathbf{u}_{M+1} =&\mathbf{C}_{R}\,\mathbf{a}_{tr} 
           \end{array}\right.,  
\end{equation} where the vector $\mathbf{a}_{in}$ is the amplitude of the right-going
incoming wave, $\mathbf{a}_{ref}$ is the amplitude of the reflected
wave, and $\mathbf{a}_{tr}$ is the amplitude of the transmitted wave.
The latter two vectors are to be calculated given the boundary conditions.
The first boundary condition is given by setting the vector $\mathbf{a}_{in}$.
This is usually taken as a unit vector, whose components are all zero,
except for the component, corresponding to one mode, which is set
to unity. The second ``boundary condition'' imposed on the wave
function is that the left-going part in the right lead vanishes.

Making use of Bloch's theorem, we have \begin{equation}     
    \left\{\begin{array}{lr}         
              \mathbf{u}_{-1}=&\mathbf{C}_{L}^{a}\,\boldsymbol{\Lambda}_{L}^{a}\mathbf{a}_{in}+
                       \mathbf{C}_{L}\,\boldsymbol{\Lambda}_{L}\mathbf{a}_{ref},\\
              \mathbf{u}_{M+2}=&\mathbf{C}_{R}\,\boldsymbol{\Lambda_{R}}\mathbf{a}_{tr}
           \end{array}\right. . 
\end{equation}

We introduce the pseudo-inverse matrices $\tilde{\mathbf{C}_{L}}$,
$\tilde{\mathbf{C}_{L}^{a}}$, and $\tilde{\mathbf{C}_{R}}$ such
that $\tilde{\mathbf{C}_{L}}\mathbf{C}_{L}=\mathbf{I}$, $\tilde{\mathbf{C}_{L}^{a}}\mathbf{C}_{L}^{a}=\mathbf{I}$,
and $\tilde{\mathbf{C}_{R}}\mathbf{C}_{R}=\mathbf{I}$, where $\mathbf{I}$
is the $n\times n$ identity matrix. Using these matrices, the wave
function component $\mathbf{u}_{-1}$ is expressed as 
\[
\mathbf{u}_{-1}=\mathbf{\mathbf{C}}_{L}^{a}\,\mathbf{\boldsymbol{\Lambda}}_{L}^{a}\,\tilde{\mathbf{C}}_{L}^{a}\,\mathbf{C}_{L}^{a}\,\mathbf{a}_{in}+\mathbf{C}_{L}\,\boldsymbol{\Lambda}_{L}\,\tilde{\mathbf{C}}_{L}\,\mathbf{C}_{L}\mathbf{a}_{ref},
\]
 and given that $\mathbf{C}_{L}\mathbf{a}_{ref}=\mathbf{u}_{0}-\mathbf{C}_{L}^{a}\,\mathbf{a}_{in}$,
we have
\[
\mathbf{u}_{-1}=\mathbf{\mathbf{C}}_{L}^{a}\,\boldsymbol{\Lambda}_{L}^{a}\,\tilde{\mathbf{C}}_{L}^{a}\,\mathbf{C}_{L}^{a}\,\mathbf{a}_{\mathrm{in}}+\mathbf{C}_{L}\,\mathbf{\mathbf{\boldsymbol{\Lambda}}}_{L}\,\tilde{\mathbf{C}}_{L}\left(\mathbf{u}_{0}-\mathbf{C}_{L}^{a}\,\mathbf{a}_{\mathrm{in}}\right).
\]
Introducing the Bloch matrices 
\begin{equation}
\mathbf{B}_{L}=\mathbf{C}_{L}\,\mathbf{\boldsymbol{\Lambda}}_{L}\,\tilde{\mathbf{C}}_{L}\label{eq:B_L_WFM}
\end{equation}
and
\begin{equation}
\mathbf{B}_{L}^{a}=\mathbf{\mathbf{C}}_{L}^{a}\,\boldsymbol{\Lambda}_{L}^{a}\,\tilde{\mathbf{C}}_{L}^{a},\label{eq:B_La_WFM}
\end{equation}
we obtain
\begin{equation}
\mathbf{u}_{-1}=\mathbf{B}_{L}^{a}\,\mathbf{C}_{L}^{a}\,\mathbf{a}_{in}+\mathbf{B}_{L}\,\mathbf{u}_{0}-\mathbf{B}_{L}\,\mathbf{C}_{L}^{a}\,\mathbf{a}_{\mathrm{in}},
\end{equation}
and finally
\begin{equation}
\mathbf{u}_{-1}=\left(\mathbf{B}_{L}^{a}-\mathbf{B}_{L}\right)\mathbf{C}_{L}^{a}\,\mathbf{a}_{\mathrm{in}}+\mathbf{B}_{L}\,\mathbf{u}_{0}.
\end{equation}
Similarly for the right lead, we introduce the Bloch matrix 
\begin{equation}
\mathbf{B}_{R}=\mathbf{C}_{R}\,\boldsymbol{\Lambda}_{R}\,\tilde{\mathbf{C}}_{R},\label{eq:B_R_WFM}
\end{equation}
 and write
\begin{equation}
\mathbf{u}_{M+2}=\mathbf{C}_{R}\,\boldsymbol{\Lambda}_{R}\,\tilde{\mathbf{C}}_{R}\,\mathbf{C}_{R}\,\mathbf{a}_{tr}=\mathbf{B}_{R}\,\mathbf{C}_{R}\,\mathbf{a}_{tr}=\mathbf{B}_{R}\,\mathbf{u}_{M+1}.
\end{equation}
Using the above, we finally obtain the following system of equations
for the central part
\[
(\omega^{2}\mathbf{I}-\boldsymbol{\mathbf{H}}_{eff}^{W})\left[\begin{array}{c}
\mathbf{u}_{0}\\
\vdots\\
\vdots\\
\mathbf{u}_{M+1}
\end{array}\right]=\left[\begin{array}{c}
\mathbf{q}_{\mathrm{inj}}\\
0\\
\vdots\\
0
\end{array}\right],
\]
with
\begin{equation}
\mathbf{H}_{eff}^{W}=\left[\begin{array}{ccccccc}
\mathbf{H}_{LS} & \mathbf{H}_{01} & 0 & 0 & 0 & 0 & 0\\
\mathbf{H}_{01}^{\dagger} & \mathbf{H}_{1} & \mathbf{H}_{12} & 0 & 0 & 0 & 0\\
0 & \mathbf{H}_{12}^{\dagger} & \mathbf{H}_{2} & \ddots & 0 & 0 & 0\\
0 & 0 & \ddots & \ddots & \ddots & 0 & 0\\
0 & 0 & 0 & \ddots & \mathbf{H}_{M-1} & \mathbf{H}_{M-1,M} & 0\\
0 & 0 & 0 & 0 & \mathbf{H}_{M-1,M}^{\dagger} & \mathbf{H}_{M} & \mathbf{H}_{M,M+1}\\
0 & 0 & 0 & 0 & 0 & \mathbf{H}_{M,M+1}^{\dagger} & \mathbf{H}_{RS}
\end{array}\right],\label{eq:heff_wfm}
\end{equation}
\[
\mathbf{H}_{LS}=\mathbf{H}_{0}+\mathbf{H}_{LL}^{\dagger}\mathbf{B}_{L},
\]
\[
\mathbf{H}_{RS}=\mathbf{H}_{M+1}+\mathbf{H}_{RR}\mathbf{B}_{R},
\]
and
\[
\mathbf{q}_{\mathrm{inj}}=\mathbf{H}_{LL}^{\dagger}\left(\mathbf{B}_{L}^{a}-\mathbf{B}_{L}\right)\mathbf{C}_{L}^{a}\mathbf{a}_{in}.
\]
The quantities
\begin{equation}
\Sigma_{R}^{W}=\mathbf{H}_{RR}\mathbf{B}_{R}\label{eq:WFM_R_self-energy} 
\end{equation}
and
\begin{equation}
\Sigma_{L}^{W}=\mathbf{H^{\dagger}}_{LL}\mathbf{B}_{L}\label{eq:WFM_L_self-energy} 
\end{equation}
are known, respectively, as the self-energies for the right and left
leads, and are formally introduced later in the AGF formalism. These are the defining
quantities that will allow us to establish a connection with the AGF
method and, consequently, the sought after equivalence. Defining $\mathbf{G}_{C}^{W}=\left[\omega^{2}\mathbf{I}-\boldsymbol{\mathbf{H}}_{eff}^{W}\right]^{-1}$,
the above system can inverted as
\begin{equation}
\left[\begin{array}{c}
\mathbf{u}_{0}\\
\vdots\\
\vdots\\
\mathbf{u}_{M+1}
\end{array}\right]=\boldsymbol{\mathbf{G}}_{C}^{W}\left[\begin{array}{c}
\mathbf{q}_{\mathrm{inj}}\\
0\\
\vdots\\
0
\end{array}\right].\label{u_from_q_inj}
\end{equation}

We remark at this point that only the three Bloch matrices ($\mathbf{B}_{L}$,
$\mathbf{B}_{L}^{a}$, and $\mathbf{B}_{R}$) we defined above are
needed to carry out all the subsequent analyses and computations.
As we will see below, these matrices can all be obtained from the
AGF approach. However, it is customary in prior literature\citep{khomyakov05,Ong2015}
to define eight such matrices, with four of them having no connection
at all to the AGF formalism. 

\subsection{Transmission}

For a particular incoming mode $m$ given by the $m^{\mathrm{th}}$
column of the matrix $\mathbf{C}_{L}^{a}$, the expression $\mathbf{u}_{M+1}=\mathbf{C}_{R}\,\mathbf{a}_{tr}$
reads:
\begin{equation}
\mathbf{C}_{R}\,\boldsymbol{\tau}_{m}=\mathbf{u}_{M+1,m}.
\end{equation}
We define the matrix $\boldsymbol{\tau}$ as composed of the columns
$\tau_{m}$($m=1,2,...,N$), where $N$ is the number of modes:
\[
\boldsymbol{\boldsymbol{\tau}}=(\boldsymbol{\tau}_{1}\,\boldsymbol{\tau}_{2}\,...\,\boldsymbol{\tau}_{N}).
\]
We generalize the earlier relation and obtain
\[
\mathbf{C}_{R}\,\boldsymbol{\tau}=(\mathbf{u}_{M+1,1}\,\mathbf{u}_{M+1,2}\,....\,\mathbf{u}_{M+1,N}).
\]
From Eq. \ref{u_from_q_inj}, we have 
\begin{equation}
\mathbf{u}_{M+1,m}=\mathbf{G}_{C}^{W}(M+1,0)\mathbf{q}_{\mathrm{inj}}.\label{eq:u_g_qinj}
\end{equation}
It is a common practice to fix $\mathbf{a}_{in}$ as having a unit
component only along the $m^{\mathrm{{th}}}$mode, with all the other
components vanishing. Instead of working with the vectors $\mathbf{a}_{in}$
and $\mathbf{q}_{inj}$, we define the respective matrices $\mathbf{A}_{in}$
and $\mathbf{Q}_{inj}$. Here $\mathbf{A}_{in}$ is just the $n\times n$
identity matrix, and as a consequence, $\mathbf{Q}_{inj}$ is given
as 
\begin{equation}
\mathbf{Q}_{\mathrm{inj}}=\mathbf{H}_{LL}^{\dagger}(\mathbf{B}_{L}^{a}-\mathbf{B}_{L})\mathbf{C}_{L}^{a}\mathbf{A}_{in}=\mathbf{H}_{LL}^{\dagger}(\mathbf{B}_{L}^{a}-\mathbf{B}_{L})\mathbf{C}_{L}^{a},\label{Q_0_expression}
\end{equation}
which follows from $\mathbf{A}_{in}$ being the identity matrix. We
then obtain
\[
\tau=\mathbf{C}_{R}^{-1}\mathbf{G}_{C}^{W}(M+1,0)\mathbf{Q}_{\mathrm{inj}}.
\]
The transmission amplitude is therefore found by normalizing with
respect to the current and reads
\begin{eqnarray*}
\mathbf{t}= & \mathbf{V}_{R}^{\nicefrac{1}{2}}\,\mathbf{C}_{R}^{-1}\,\mathbf{G}_{C\left(M+1,0\right)}^{W}\,\mathbf{Q}_{\mathrm{inj}}\,\left(\mathbf{V}_{L}^{a}\right)^{\nicefrac{-1}{2}}.
\end{eqnarray*}
Replacing $\mathbf{Q}_{\mathrm{inj}}$ , defined in Eq. (eq. \ref{Q_0_expression}),
with its alternative expression, $\mathbf{Q}_{\mathrm{inj}}=i\left(\mathbf{C}_{L}^{a\dagger}\right)^{-1}\mathbf{V}_{L}^{a}$(\textit{\textcolor{black}{cf.}}
\ref{app:injection-matrix}), we arrive at the final expression for
the transmission amplitude:
\begin{equation}
\mathbf{t}=i\,\mathbf{V}_{R}^{\nicefrac{1}{2}}\,\mathbf{C}_{R}^{-1}\,\mathbf{G}_{C\left(M+1,0\right)}^{W}\left(\mathbf{C}_{L}^{a\dagger}\right)^{-1}\left(\mathbf{V}_{L}^{a}\right)^{\nicefrac{1}{2}}.\label{t_wfm}
\end{equation}
This is similar to the expression arrived at by Komyakov \textit{et
al.}\citep{khomyakov05}, starting from the AGF formalism. The difference
lies in the way they define the Bloch matrix describing the incoming
wave. As a result, in their case, even the evanescent modes yield
non-vanishing elements in $\mathbf{Q}_{\mathrm{inj}}$, which they
then proceed to eliminate ``by hand'', by considering the group
velocities. The evanescent modes, having a vanishing group velocity,
the corresponding entries in $\mathbf{Q}_{\mathrm{inj}}$ are set
to zero. However, in our case the evanescent modes are eliminated
``naturally'' thanks to the clever choice of the modes entering
our definition of the Bloch matrix $\mathbf{B}_{L}^{a}$. By naturally
we mean that there is no need to scan over the modes is search of
the decaying ones and then zero out the corresponding entries in $\mathbf{Q}_{inj}$.We
must indicate that Komyakov \textit{et al.}\citep{khomyakov05} define
a matrix $\mathbf{Q}_{0}$ (their eq. 14), which when multiplied by
their $\mathbf{U}_{L}(+)$ (their eq. 55) on the right, yields a matrix
$\mathbf{Q}_{\mathrm{inj}}^{\mathrm{K}}$, to which we compare our
$\mathbf{Q}_{\mathrm{inj}}$.

\section{AGF formalism}

\subsection{A brief review of the conventional AGF formalism}

We now provide a brief overview of the well-known AGF method with
the aim to demonstrate its equivalence the WFM formalism described
earlier. And more importantly, once the complete equivalence is shown,
we will exploit it to extract the individual, i.e. mode-resolved transmission
amplitudes. For a more detailed treatment of the AGF method, we refer
the reader to prior literature \citep{Mingo2009,sadasivan2014}.

In the single particle formalism the retarded Green's function $\mathbf{G}$
, also known as the resolvent, is defined as the operator inverse
of the Hamiltonian $\mathbf{H}$ (given in Eq. \ref{006}):
\[
(\omega^{2}\mathbf{I}+i\eta-\mathbf{H})\mathbf{G}(\omega,\mathbf{k}_{\perp})=\mathbf{I},
\]
where $\eta$ is a positive infinitesimal. A similar equation with
$-i\eta$ (instead of $+i\eta$) defines the advanced Green's function
$\mathbf{G}^{a}$, which can also be viewed as the conjugate of the
retarded function $\mathbf{G}$. The element $\mathbf{G}_{ij}$ represents
the response of slice $j$ to a small vibration of slice $i$. It
is therefore related to the transmission probability between slices
$i$ and $j$. In matrix form, the above equation reads
\[
\left[\begin{array}{ccc}
\omega^{2}\mathbf{I}-\mathbf{H}_{L} & -\mathbf{H}_{LL} & 0\\
-\mathbf{H}_{LL}^{\dagger} & \omega^{2}\mathbf{I}-\mathbf{H}_{C} & -\mathbf{H}_{RR}\\
0 & -\mathbf{H}_{RR}^{\dagger} & \omega^{2}\mathbf{I}-\mathbf{H}_{R}
\end{array}\right]\left[\begin{array}{ccc}
\mathbf{G}_{L} & \mathbf{G}_{LC} & \mathbf{G}_{LR}\\
\mathbf{G}_{CL} & \mathbf{G}_{C} & \mathbf{G}_{CR}\\
\mathbf{G}_{RL} & \mathbf{G}_{RC} & \mathbf{G}_{R}
\end{array}\right]=\left[\begin{array}{ccc}
\mathbf{I} & 0 & 0\\
0 & \mathbf{I} & 0\\
0 & 0 & \mathbf{I}
\end{array}\right].
\]
We can then solve for the central part alone and obtain
\[
\mathbf{G}_{C}=\left[\omega^{2}\mathbf{I}-\mathbf{H}_{C}-\mathbf{\boldsymbol{\Sigma}}_{L}-\boldsymbol{\Sigma}_{R}\right]^{-1},
\]
where $\boldsymbol{\Sigma}_{L}$ and $\boldsymbol{\Sigma}_{R}$ are
the self-energies containing the effects due to connecting the central
region to the left and right leads, respectively. They are given by
\begin{equation}
\boldsymbol{\Sigma}_{L}=\mathbf{H}_{LL}^{\dagger}\,\mathbf{g}_{L}\,\mathbf{H}_{LL},\label{eq:SigmaL_gL}
\end{equation}
and
\begin{equation}
\mathbf{\boldsymbol{\Sigma}}_{R}=\mathbf{H}_{RR}\,\mathbf{g}_{R}\,\mathbf{H}_{RR}^{\dagger},\label{eq:SigmaR_gR}
\end{equation}
where $\mathbf{g}_{L}$ and $\mathbf{g}_{R}$ are the surface Green's
functions of the left and right leads, respectively\citep{Mingo2009,sadasivan2014}.
In this work they are computed using a numerically efficient algorithm
devised by Lopez-Sancho\citep{Sancho1985,velev2004,Zhu2016}. The
central part is as such represented by an effective Hamiltonian
\begin{equation}
\mathbf{H}_{eff}^{G}=\left[\begin{array}{ccccccc}
\mathbf{H}_{0}+\boldsymbol{\Sigma}_{L} & \mathbf{H}_{01} & 0 & 0 & 0 & 0 & 0\\
\mathbf{H}_{01}^{\dagger} & \mathbf{H}_{1} & \mathbf{H}_{12} & 0 & 0 & 0 & 0\\
0 & \mathbf{H}_{12}^{\dagger} & \mathbf{H}_{2} & \ddots & 0 & 0 & 0\\
0 & 0 & \ddots & \ddots & \ddots & 0 & 0\\
0 & 0 & 0 & \ddots & \mathbf{H}_{M-1} & \mathbf{H}_{M-1,M} & 0\\
0 & 0 & 0 & 0 & \mathbf{H}_{M-1,M}^{\dagger} & \mathbf{H}_{M} & \mathbf{H}_{M,M+1}\\
0 & 0 & 0 & 0 & 0 & \mathbf{H}_{M,M+1}^{\dagger} & \mathbf{H}_{M+1}+\boldsymbol{\Sigma}_{R}
\end{array}\right],\label{heff_agf}
\end{equation}
such that
\[
(\omega^{2}\mathbf{I}-\mathbf{H}_{eff}^{G})\mathbf{G}_{C}=\mathbf{I}.
\]
The device Green's functions $\mathbf{G}_{C}$ and $\mathbf{G}_{C}^{a}$
are then used to compute the total transmission via the Caroli formula
\citep{caroli71}:
\begin{equation}
T=\mathrm{trace}\left(\mathbf{G_{\mathrm{C}}}\,\mathbf{\boldsymbol{\Gamma}}_{L}\,\mathbf{G_{\mathrm{C}}^{\mathit{a}}}\,\mathbf{\boldsymbol{\Gamma}}_{R}\right),\label{eq:caroli_formula}
\end{equation}
where
\[
\mathbf{\boldsymbol{\Gamma}}_{R(L)}=i\left(\mathbf{\boldsymbol{\Sigma}}_{R(L)}-\mathbf{\boldsymbol{\Sigma}}_{R(L)}^{\dagger}\right).
\]

\subsection{Equivalence of WFM and AGF}

The effective Hamiltonian $\mathbf{H}_{eff}^{G}$ defined in Eq. \ref{heff_agf}
is to be compared to the effective Hamiltonian $\mathbf{H}_{eff}^{W}$
obtained in the WFM formalism (Eq. \ref{eq:heff_wfm}). The two must
indeed be identical. As a result, we can now proceed to identifying
the elements appearing in the two expressions of the effective Hamiltonian.
The right and left self-energies pertaining to the effects of the
right and left leads, respectively, in the WFM formalism are given
in Eqs. \ref{eq:WFM_R_self-energy} and \ref{eq:WFM_L_self-energy}. The equivalent quantities given
in the AGF formalism are written in Eqs. \ref{eq:SigmaL_gL} and \ref{eq:SigmaR_gR}.
Equating these quantities term by term leads to the following equivalences
\begin{eqnarray}
\mathbf{B}_{R} & = & \mathbf{g}_{R}\,\mathbf{H}_{RR}^{\dagger},\label{appa-2-2}\\
\mathbf{B}_{L} & = & \mathbf{g}_{L}\,\mathbf{H}_{LL}.\label{appb-2-2}
\end{eqnarray}
The surface Green's functions are then related to the Bloch matrices
as
\begin{eqnarray}
\mathbf{g}_{R} & = & \mathbf{B}_{R}\,\left(\mathbf{H}_{RR}^{\dagger}\right)^{-1},\label{appa-2-2-1}\\
\mathbf{g}_{L} & = & \mathbf{B}_{L}\,\left(\mathbf{H}_{LL}\right)^{-1},\label{appb-2-2-1}
\end{eqnarray}
for the right and left leads, respectively. Similarly, for the advanced
quantities, the self-energies are given in the WFM formalism as
\begin{eqnarray}
\mathbf{\boldsymbol{\Sigma}}_{R}^{a(W)} & = & \mathbf{H}_{RR}\,\mathbf{B}_{R}^{a},\label{appa-2-3}\\
\mathbf{\boldsymbol{\Sigma}}_{L}^{a(W)} & = & \mathbf{H}_{LL}^{\dagger}\,\mathbf{B}_{L}^{a},\label{appb-2-3}
\end{eqnarray}
and in the AGF formalism as
\begin{eqnarray}
\mathbf{\boldsymbol{\Sigma}}_{R}^{a} & = & \mathbf{H}_{RR}\,\mathbf{g}_{R}^{a}\,\mathbf{H}_{RR}^{\dagger},\label{appa-2-1-1}\\
\mathbf{\boldsymbol{\Sigma}}_{L}^{a} & = & \mathbf{H}_{LL}^{\dagger}\,\mathbf{g}_{L}^{a}\,\mathbf{H}_{LL}.\label{appb-2-1-1}
\end{eqnarray}
Equating these quantities leads to the following equivalences
\begin{eqnarray}
\mathbf{B}_{R}^{a} & = & \mathbf{g}_{R}^{a}\,\mathbf{H}_{RR}^{\dagger},\label{appa-2-2-2}\\
\mathbf{B}_{L}^{a} & = & \mathbf{g}_{L}^{a}\,\mathbf{H}_{LL}.\label{appb-2-2-2}
\end{eqnarray}
The advanced surface GFs are then related to the advanced Bloch matrices
as
\begin{eqnarray}
\mathbf{g}_{R}^{a} & = & \mathbf{B}_{R}^{a}\left(\mathbf{H}_{RR}^{\dagger}\right)^{-1},\label{appa-2-2-1-1}\\
\mathbf{g}_{L}^{a} & = & \mathbf{B}_{L}^{a}\left(\mathbf{H}_{LL}\right)^{-1}.\label{appb-2-2-1-1}
\end{eqnarray}
Finally, we arrive at the following expressions for the normalized
group velocity matrices needed in the expression of the transmission
amplitudes:
\begin{equation}
\begin{array}{ccc}
\mathbf{V}_{R} & = & \mathbf{C}_{R}^{\dagger}\,\Gamma_{R}\,\mathbf{C}_{R,}\\
\mathbf{V}_{L}^{a} & = & \mathbf{C}_{L}^{a\dagger}\,\Gamma_{L}\,\mathbf{C}_{L}^{a}.
\end{array}\label{vr_c_gamma}
\end{equation}
with the details given in \ref{app:velocity-matrices}. These are
identical to the expressions found by\textcolor{black}{{} Komyakov }\textit{\textcolor{black}{et
al.}}\textcolor{black}{\citep{khomyakov04} }who used a rather complicated
route. We note that these are the key ingredients to resolving, in
the next sub-section, the total transmission function in terms of
individual phonon contributions. But, before moving on, we think it
is important to keep track of the most important quantities introduced
so far within the two approaches. These are shown in Table \ref{tab:A-comp-wfm-agf-quant}
and in the order that they are computed in each method.
\begin{table}
\begin{centering}
\begin{tabular}{|c|c|}
\hline 
WFM & AGF\tabularnewline
\hline 
\hline 
$H_{L}$, $H_{R}$ & $H_{L}$, $H_{R}$\tabularnewline
\hline 
$B_{L}$, $B_{L}^{a}$, $B_{R}$ & $g_{L}$, $g_{R}$\tabularnewline
\hline 
$\Sigma_{L}^{W}$,$\Sigma_{R}^{W}$ & $\Sigma_{L}$,$\Sigma_{R}$\tabularnewline
\hline 
$H_{C}^{W}$ & $H_{C}$\tabularnewline
\hline 
$\left(H_{C}^{W}\right)^{-1}\equiv G_{C}^{W}$ & $H_{C}^{-1}\equiv G_{C}$\tabularnewline
\hline 
$T$ & $T$\tabularnewline
\hline 
\end{tabular}
\par\end{centering}
\caption{A comparative list of the quantities defined in the two formalisms,
WFM and AGF, and in the same order of their calculations.\label{tab:A-comp-wfm-agf-quant}}
\end{table}

\subsection{Mode-resolved transmissions}

As mentioned earlier, a number of attempts were made aiming at deriving
expressions\citep{khomyakov05,Huang2011,Sadasivan2017,Ong2015,Ong2018,Wimmer2009}
for the mode-resolved transmission probabilities. However, these reports
were based on involved mathematical derivations and it is not yet
obvious whether all agree in the end. In our case, we will exploit
the identities in Eq. \ref{vr_c_gamma} for a straightforward derivation
of the desired relation.One advantage of our derivation is that it
yields to a straightforward physical interpretation.

Indeed, we can obtain $\Gamma_{R}$ and $\Gamma_{L}$ in terms of
the normalized group velocity matrices form Eq. \ref{vr_c_gamma}
as:
\begin{equation}
\mathbf{\boldsymbol{\Gamma}}_{R}=\left(\mathbf{C}_{R}^{\dagger}\right)^{-1}\,\mathbf{V}_{R}\,\left(\mathbf{C}_{R}\right)^{-1}\label{gamma_r}
\end{equation}
and
\begin{equation}
\mathbf{\boldsymbol{\Gamma}}_{L}=\left(\mathbf{C}_{L}^{a\dagger}\right)^{-1}\,\mathbf{V}_{L}^{a}\,\left(\mathbf{C}_{L}^{a}\right)^{-1}.\label{gamma_l}
\end{equation}
Next, we define the following operators:
\begin{equation}
\boldsymbol{\Pi}_{R}=\left(\mathbf{C}_{R}^{\dagger}\right)^{-1}\,\left(\mathbf{V}_{R}\right)^{\nicefrac{1}{2}}\label{Pi_r}
\end{equation}
and
\begin{equation}
\boldsymbol{\Pi}_{L}=\left(\mathbf{C}_{L}^{a\dagger}\right)^{-1}\,\left(\mathbf{V}_{L}^{a}\right)^{\nicefrac{1}{2}},\label{Pi_l}
\end{equation}
such that
\[
\boldsymbol{\Pi}_{R}\,\boldsymbol{\Pi}_{R}^{\dagger}=\left(\mathbf{C}_{R}^{\dagger}\right)^{-1}\,\mathbf{V}_{R}\,\left(\mathbf{C}_{R}\right)^{-1}=\mathbf{\boldsymbol{\Gamma}}_{R},
\]
and
\[
\boldsymbol{\Pi}_{L}\,\boldsymbol{\Pi}_{L}^{\dagger}=\left(\mathbf{C}_{L}^{a\dagger}\right)^{-1}\,\mathbf{V}_{L}^{a}\,\left(\mathbf{C}_{L}^{a}\right)^{-1}=\mathbf{\boldsymbol{\Gamma}}_{L}.
\]
The Caroli formula (Eq. \ref{eq:caroli_formula}) can now be recast
as
\begin{eqnarray}
T= &  & \mathrm{trace}\left(\mathbf{G}^{r}\,\mathbf{\boldsymbol{\Gamma}}_{L}\,\mathbf{G}^{a}\,\mathbf{\boldsymbol{\Gamma}}_{R}\right)\nonumber \\
= &  & \mathrm{trace}\left(\mathbf{G}^{r}\,\boldsymbol{\Pi}_{L}\,\boldsymbol{\Pi}_{L}^{\dagger}\,\mathbf{G}^{a}\,\boldsymbol{\Pi}_{R}\,\boldsymbol{\Pi}_{R}^{\dagger}\right)\nonumber \\
= &  & \mathrm{trace}\left(\boldsymbol{\Pi}_{R}^{\dagger}\,\mathbf{G}^{r}\,\boldsymbol{\Pi}_{L}\,\boldsymbol{\Pi}_{L}^{\dagger}\,\mathbf{G}^{a}\,\boldsymbol{\Pi}_{R}\right)\nonumber \\
= &  & \mathrm{trace}\left(\mathbf{t}\,\mathbf{t}^{\dagger}\right),
\end{eqnarray}
where we made use of the cyclic property of the trace. We can identify
the matrices appearing in the last line of the equality as
\begin{equation}
\mathbf{t}=i\boldsymbol{\,\Pi}_{R}^{\dagger}\,\mathbf{G}_{C}\,\boldsymbol{\Pi}_{L}=i\;\left(\mathbf{V}_{R}\right)^{\nicefrac{1}{2}}\,\left(\mathbf{C}_{R}\right)^{-1}\mathbf{G}_{C}\,\left(\mathbf{C}_{L}^{a\dagger}\right)^{-1}\,\left(\mathbf{V}_{L}^{a}\right)^{\nicefrac{1}{2}},\label{t_agf}
\end{equation}
and
\[
\mathbf{t}^{\dagger}=-i\boldsymbol{\,\Pi}_{L}^{\dagger}\,\mathbf{G}_{C}^{a}\,\boldsymbol{\Pi}_{R}=-i\;\left(\mathbf{V}_{L}^{a}\right)^{\nicefrac{1}{2}}\,\left(\mathbf{C}_{L}^{a}\right)^{-1}\mathbf{G}_{C}^{a}\,\left(\mathbf{C}_{R}^{\dagger}\right)^{-1}\,\left(\mathbf{V}_{R}\right)^{\nicefrac{1}{2}}.
\]
We note that the matrices $\mathbf{C}_{R}$ and $\mathbf{C}_{L}^{a}$
can be readily obtained from the Green's functions as shown in \ref{app:eigenvector-matrices}.
We can clearly see that the transmission matrix given by Eq. \ref{t_agf}
is identical to its definition in the WFM formalism (Eq. \ref{t_wfm}),
with the identity $\boldsymbol{\mathbf{G}}_{C}^{W}\equiv\boldsymbol{\mathbf{G}}_{C}$.
The mode-resolved transmissions, say from mode $m$ in the left lead
to mode $k$ in the right lead, are given as the matrix elements
\begin{equation}
t_{k,m}=\left\{ i\,\mathbf{V}_{R}^{\nicefrac{1}{2}}\,\mathbf{C}_{R}^{-1}\,\mathbf{G}_{C}\left(\mathbf{C}_{L}^{a\dagger}\right)^{-1}\,\left(\mathbf{V}_{L}^{a}\right)^{\nicefrac{1}{2}}\right\} _{k,m}.\label{eq:matrix_elmts_t}
\end{equation}
This expression gives the correct mode-resolved transmission functions,
but one has to perform all the matrix operations involved in obtaining
$t$. It would be convenient to derive a simpler expression where
matrix elements could be computed without the unnecessary computational
overhead of these matrix operations. In order to derive the desired
expression, we multiply both sides of the identities given in Eqs.
\ref{gamma_r} and \ref{gamma_l}, by $\mathbf{V_{\mathit{R}}^{\mathrm{\mathit{\mathrm{-1/2}}}}C}_{R}^{\dagger}$
on the right and by $\boldsymbol{C}_{L}^{a}\left(\mathbf{V}_{L}^{a}\right)^{\mathrm{-1/2}}$
on the left, respectively, and obtain\textcolor{black}{
\[
\begin{array}{ccc}
\mathbf{V_{\mathit{R}}^{\mathrm{\mathit{\mathbf{\mathrm{-1/2}}}}}C}_{R}^{\dagger}\,\boldsymbol{\Gamma}_{R} & = & \mathbf{V^{\mathit{1/2}}}_{R}\,\mathbf{C}_{R}^{-1},\\
\boldsymbol{\Gamma}_{L}\,\mathbf{C}_{L}^{a}\left(\mathbf{V}_{L}^{a}\right)^{\mathbf{\mathrm{-1/2}}} & = & \left(\mathbf{C}_{L}^{a\dagger}\right)^{-1}\,\left(\mathbf{V}_{L}^{a}\right)^{\mathbf{\mathrm{1/2}}}.
\end{array}
\]
Using} these identities, we can, finally, rewrite the transmission
matrix (Eq. \ref{t_agf}) in a useful form as
\begin{equation}
\mathbf{t}=\mathrm{i}\,\mathbf{V}_{R}^{\nicefrac{-1}{2}}\,\mathbf{C}_{R}^{\dagger}\,\boldsymbol{\Gamma}_{R}\,\boldsymbol{G}_{C}\boldsymbol{\Gamma}_{L}\,\mathbf{C}_{L}^{a}\,\left(\mathbf{V}_{L}^{a}\right)^{\nicefrac{-1}{2}}.\label{eq:t_agf}
\end{equation}
Since the above expression contains the matrices $\mathbf{C}_{R}$
and $\mathbf{C}_{L}^{a}$ instead of their inverses, we can extract
straightaway the mode-resolved transmissions as
\begin{equation}
t_{k,m}=\frac{\mathit{i}}{\sqrt{v_{R,k}\,v_{L,m}^{a}}}\,\mathbf{u}_{R,k}^{\dagger}\,\boldsymbol{\Gamma}_{R}\,\mathbf{G}_{C}\boldsymbol{\Gamma}_{L}\,\mathbf{u}_{L,m}^{a}.\label{eq:mode_resolved_agf}
\end{equation}
Although similar expressions have been derived in previous works\citep{khomyakov05,Sanvito1999,Wimmer2009,zhang2017},
by means of mathematically sound manipulations, our derivation is
much simpler and, moreover, the final expression has a very straightforward
physical interpretation. Indeed, reading the expression from right
to left, we have the complete picture of a phonon travelling from
the left to the right leads, through the device region: $\mathbf{u}_{L,m}^{a}$
is the incoming wave or phonon in the mode $m$ from the left lead,
$\boldsymbol{\Gamma}_{L}$ gives the part of the incoming wave that
is transmitted into the device, $\mathbf{G}_{C}$ is the propagator
that ``transports'' that part to the other side of the device, $\boldsymbol{\Gamma}_{R}$
gives the part that is then transmitted to the right lead, and finally,
the multiplication by $\mathbf{u}_{R,k}^{\dagger}$ is merely a projection
that yields the probability amplitude to end up in the mode $k$ of
the right lead. We note that the group velocities serve only as normalizing
factors for the eigenvectors $\mathbf{u}_{L,m}^{a}$ and $\mathbf{u}_{R,k}^{\dagger}$.
However, unlike in Eq. \ref{eq:matrix_elmts_t}, care must be taken
when using Eq. \ref{eq:mode_resolved_agf}, as a result of the presence
of the group velocities in the denominator. It is to be understood
that when either of the modes ($m$ or $k$) is evanescent, the corresponding
$t_{k,m}$ vanishes, and no calculation is needed. The above expression
is to be compared to the similar formula, obtained earlier using the
WFM formalism (Eq. \ref{t_wfm}), which, in hindsight and using the
equivalence of the two formalisms, can readily be transformed into
Eq. \ref{eq:t_agf}. Our formula (Eq. \ref{eq:mode_resolved_agf})
is also to be compared to the Fisher-Lee expression, which is valid
only if the eigenvectors are orthogonal. Our result is therefore more
general than that of Fisher-Lee which has been, for years, the text-book
expression\citep{datta02,Ferry2001}. 

The question of whether the two methods are equivalent has remained
unsolved. The reason is that it has been widely held that the AGF
includes both evanescent and propagating modes, whereas in the WFM
the first are discarded and hence do not contribute to the transmission
function\citep{krstic02,zhang2017}. We do, however, disagree with
this analysis and assert that even in the AGF method the evanescent
modes are implicitly discarded through the ``escape rate'' matrices.
Indeed, in Eq. \ref{eq:mode_resolved_agf} the term $\boldsymbol{\Gamma}_{L}\,\mathbf{u}_{L,m}^{a}$
, we refer to as the injection vector in the WFM formalism (Eq. \ref{eq:Qinj_appendix}),
selects out the evanescent modes, as can be seen below (next section)
where we compute this term.

\begin{figure}
\begin{centering}
\includegraphics[scale=0.25]{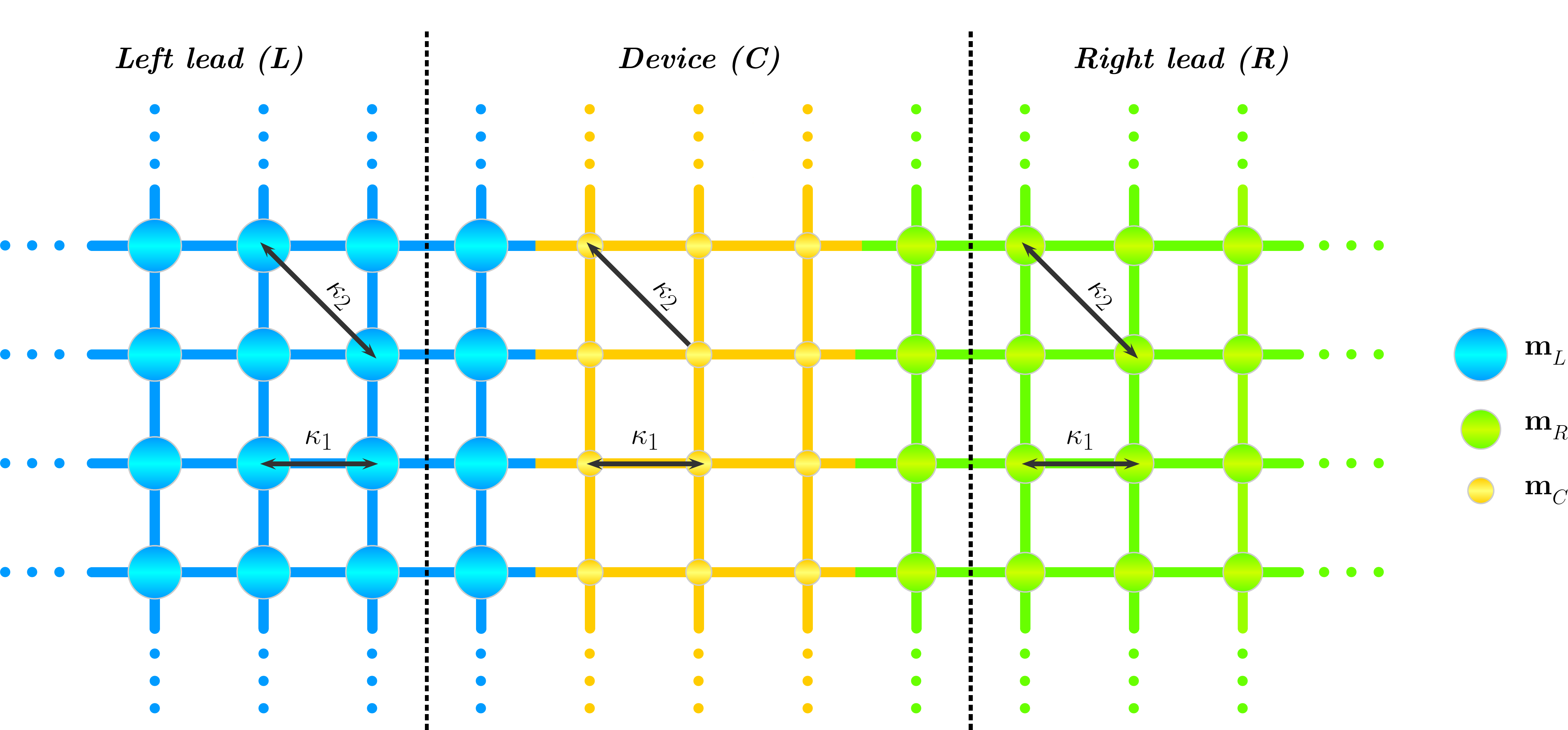}
\par\end{centering}
\caption{A schematic representation the system consisting of a square lattice
with the same lattice constant and force constants ($\kappa_{1}$
for first-neighbor and $\kappa_{2}$ for second-neighbor) throughout.
The masses are different in the three regions: they are $m_{L}$ in
the left lead, $m_{C}$ in the central region, and $m_{R}$ in the
right lead\label{fig:square_lattice_model}.}
\end{figure}

\begin{table}
\begin{centering}
\begin{tabular}{|c|c|c|}
\hline 
 & $\omega=1.55$ & $\omega=2.20$\tabularnewline
\hline 
\hline 
$\triangle\mathbf{g}_{L}$ & $\left[\begin{array}{cc}
2.1156 & 8.4644\\
8.4688 & 30.4458
\end{array}\right]$$10^{-12}$ & $\left[\begin{array}{cc}
2.1717 & 0.7383\\
0.7402 & 0.3862
\end{array}\right]$$10^{-12}$\tabularnewline
\hline 
$\triangle\mathbf{g}_{R}$ & $\left[\begin{array}{cc}
0.9315 & 3.2897\\
3.2895 & 12.0360
\end{array}\right]$$10^{-12}$ & $\left[\begin{array}{cc}
0.7092 & 0.9069\\
0.9066 & 2.3428
\end{array}\right]$$10^{-12}$\tabularnewline
\hline 
$\triangle\mathbf{B}_{L}$ & $\left[\begin{array}{cc}
2.0924 & 3.0534\\
8.9079 & 11.0897
\end{array}\right]$$10^{-12}$ & $\left[\begin{array}{cc}
2.5303 & 1.0733\\
1.1024 & 0.2075
\end{array}\right]$$10^{-12}$\tabularnewline
\hline 
$\triangle\mathbf{B}_{R}$ & $\left[\begin{array}{cc}
2.3381 & 1.8484\,\;\,\\
7.8625 & 6.8557
\end{array}\right]$$10^{-12}$ & $\left[\begin{array}{cc}
1.5516 & 0.6837\\
2.4664 & 1.2571
\end{array}\right]$$10^{-12}$\tabularnewline
\hline 
$\triangle\boldsymbol{\Gamma}_{L}$ & $\left[\begin{array}{cc}
3.0993 & 6.0824\,\;\,\\
6.0837 & 3.0986
\end{array}\right]$$10^{-12}$ & $\left[\begin{array}{cc}
3.3946 & 2.0217\\
2.0255 & 0.8398
\end{array}\right]$$10^{-12}$\tabularnewline
\hline 
$\triangle\boldsymbol{\Gamma}_{R}$ & $\left[\begin{array}{cc}
3.1228 & 9.1446\,\;\,\\
9.1450 & 3.1148
\end{array}\right]$$10^{-12}$ & $\left[\begin{array}{cc}
5.1781 & 3.0598\\
3.0615 & 0.6453
\end{array}\right]$$10^{-12}$\tabularnewline
\hline 
$\triangle\mathbf{V}_{L}^{a}$ & $\left[\begin{array}{cc}
3.0985 & 5.1429\,\;\,\\
5.1422 & 3.1002
\end{array}\right]$$10^{-12}$ & $\left[\begin{array}{cc}
0.2061 & 0.5931\\
0.5897 & 3.2494
\end{array}\right]$$10^{-12}$\tabularnewline
\hline 
$\triangle\mathbf{V}_{R}$ & $\left[\begin{array}{cc}
3.1173 & 6.5236\,\;\,\\
6.5243 & 3.1151
\end{array}\right]$$10^{-12}$ & $\left[\begin{array}{cc}
1.6979 & 1.6577\\
1.6571 & 3.1965
\end{array}\right]$$10^{-12}$\tabularnewline
\hline 
$\triangle T$ & $7.9869\times10^{-12}$ & $0.7000\times10^{-12}$\tabularnewline
\hline 
\end{tabular}
\par\end{centering}
\caption{The absolute value of the differences between various quantities computed
using the WFM and the AGF formalisms for $k_{y}=1$. The quantities
are the left ($\mathbf{g}_{L}$) and right ($\mathbf{g}_{R}$) surface
Green's functions, the left ($\mathbf{B}_{L}$) and right ($\mathbf{B}_{R}$)
Bloch matrices, the left ($\boldsymbol{\Gamma}_{L}$) and right ($\boldsymbol{\Gamma}_{R}$)
escape rates and the transmission $T$. In the AGF formalism we used
$\eta=10^{^{-12}}$.\label{tab:diffs_wfm_agf}}
\end{table}
\begin{table}
\begin{centering}
\begin{tabular}{|c|c|c|}
\hline 
 & $\mathbf{Q}_{inj}^ {}$ & $\mathbf{Q}_{inj}^{K}$\tabularnewline
\hline 
\hline 
$\omega=0.50$ & $\left[\begin{array}{cc}
0 & 0\\
0 & 0
\end{array}\right]$ & $\left[\begin{array}{cc}
-0.7730-0.0022i & -0.0017-0.5730i\\
-0.0018+0.6537i & -1.0162+0.0030i
\end{array}\right]$\tabularnewline
\hline 
$\omega=1.00$ & $\left[\begin{array}{cc}
-0.0072-1.1453i & -0.0000-0.0000i\\
-0.6058+0.0038i & 0.0000+0.0000i
\end{array}\right]$ & $\left[\begin{array}{cc}
-0.9992-0.5597i & -0.4055-0.6913i\\
-0.2960+0.5285i & -0.3641-0.6208i
\end{array}\right]$\tabularnewline
\hline 
$\omega=1.55$ & $\left[\begin{array}{cc}
0.2629-0.1315i & -2.0993+0.8788i\\
-0.2748+0.1374i & -0.1968+0.0824i
\end{array}\right]$ & $\left[\begin{array}{cc}
0.2914-0.0385i & -1.2456-1.9046i\\
-0.3046+0.0403i & -0.1168-0.1785i
\end{array}\right]$\tabularnewline
\hline 
$\omega=2.20$ & $\left[\begin{array}{cc}
-0.0000-0.0000i & -1.4994-0.5448i\\
-0.0000-0.0000i & -0.2194+0.6038i
\end{array}\right]$ & $\left[\begin{array}{cc}
0.6107+0.0604i & 1.5771-0.2403i\\
-2.4851-0.2459i & -0.0968-0.6351i
\end{array}\right]$\tabularnewline
\hline 
\end{tabular}
\par\end{centering}
\caption{The injection matrices computed with our WFM method($\mathbf{Q}_{inj}$)
and with the method used in Komyakov \textit{et al.} \citep{khomyakov05}
($\mathbf{Q}_{inj}^{K}$), for $k_{y}=1$. We consider three frequencies:$\omega=0.50$,
$\omega=1.00$, $\omega=1.55$ and $\omega=2.20$.\label{tab:q_inj_ky_1}}
\end{table}
\begin{figure}
\begin{centering}
\includegraphics[scale=0.5]{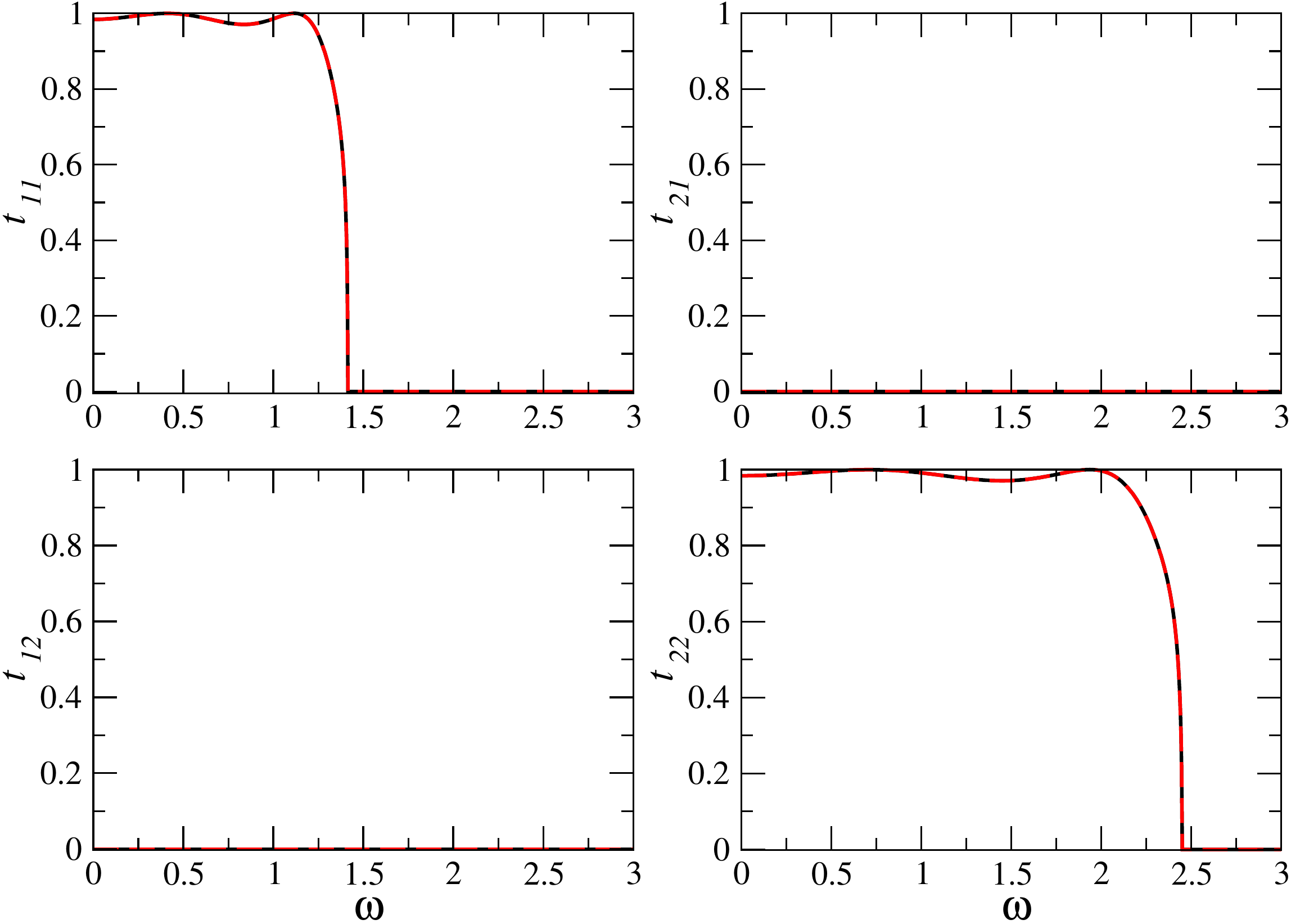}
\par\end{centering}
\caption{Mode-resolved transmission amplitudes for $k_{y}=0$. Our results
are shown in black continued line, and the results obtained using
the Fisher-Lee formula are shown in red dashed line.\label{t_km_ky_0}}
\end{figure}

\begin{figure}
\begin{centering}
\includegraphics[scale=0.5]{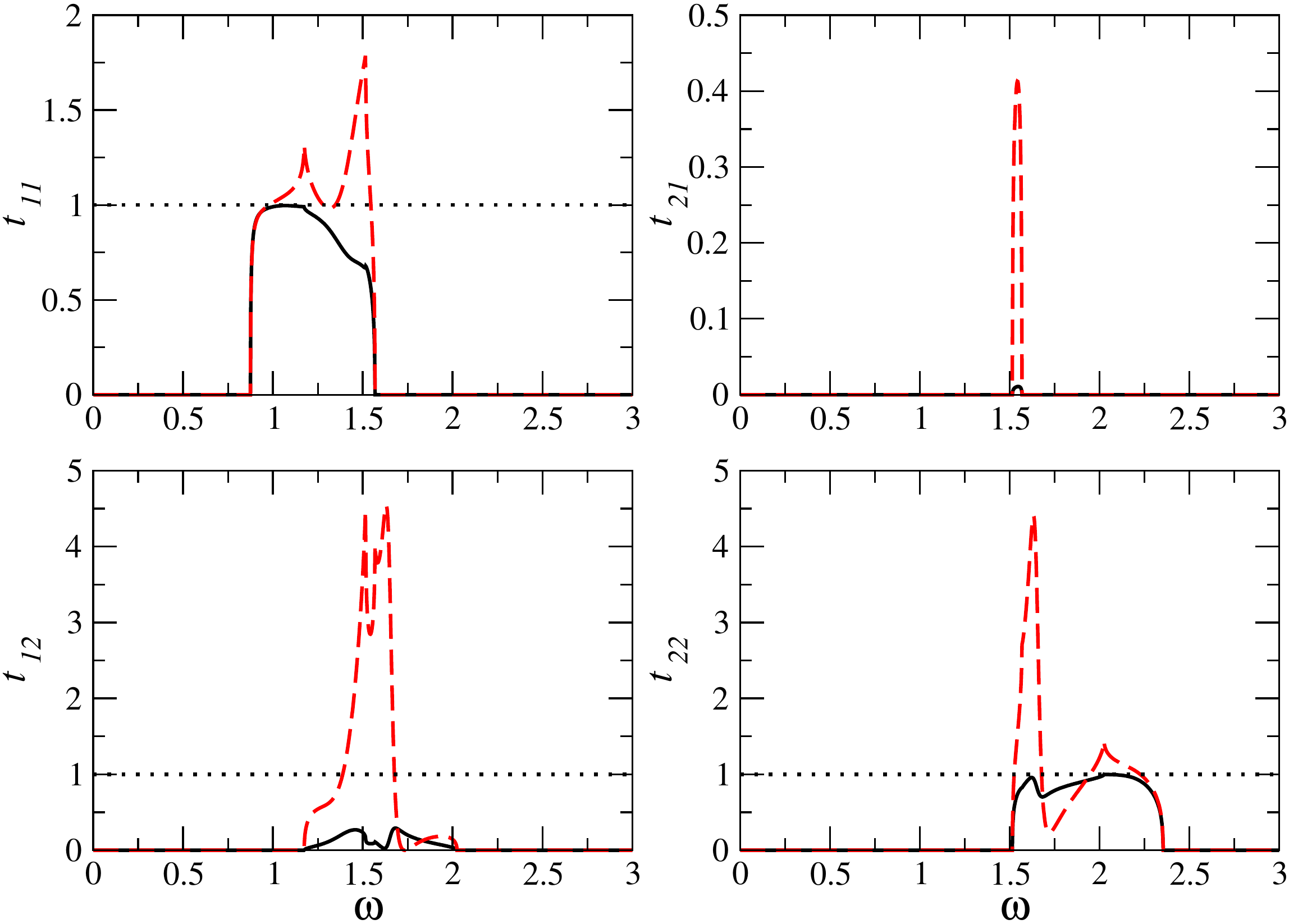}
\par\end{centering}
\caption{Mode-resolved transmission amplitudes for $k_{y}=1$. Our results
are shown in black continued line, and the results obtained using
the Fisher-Lee formula are shown in red dashed line. Where relevant
a horizontal dotted line is added to mark the maximum allowed value
of unity for the transmission amplitude.\label{t_km_ky_1}}
\end{figure}

\section{Applications}

In this part, we demonstrate the strength of our method by illustrating
its application with a simple model, and also by comparing it to previous
common methods. We consider the case of the 2D square lattice, of
lattice constant $a$, with the central part made up of three atomic
planes as shown in Fig \ref{fig:square_lattice_model}. The force
constants are identical throughout the system, but we choose to use
a different mass for each part of the device. The force constants
are $\kappa_{1}$ and $\kappa_{2}=0.5\;\kappa_{1}$ for the first
and second-neighbor interactions, respectively. The masses are $m_{L}$
for the left lead, $m_{R}=0.6\;m_{L}$ for the right lead, and $m_{C}=0.8\;m_{L}$
for the central part. We define the frequency unit $\omega_{0}=\sqrt{\kappa_{1}/m_{L}}$.
Henceforth the frequency $\omega$ is given in units of $\omega_{0}$. 

In Table \ref{tab:diffs_wfm_agf} we report the differences between
various quantities computed both with our WFM method and with the
AGF method for a wave vector $k_{y}=1$ and for two frequencies: $\omega=1.55$
and $\omega=2.20$. The latter are chosen so as to include both propagating
and evanescent or decaying modes. In the AGF method we chose the parameter
$\eta=10^{-12}$. In theory the parameter $\eta$ is infinitesimal
and one is to take the limit $\eta\to0^{+}$ to recover the correct
result. In the current calculation, however, we have no access to
a closed-form final formula, and if a limit is to be taken, it has
to be numerical in nature. Our choice of a fixed value is only meant
to serve as an illustration. It is clear from Table \ref{tab:diffs_wfm_agf}
that indeed all the relevant quantities agree perfectly when computed
with either of the methods. The differences are, indeed, due only
to the size of the parameter $\eta$. This is a clear demonstration
of the complete equivalence of our WFM method and the AGF method as
shown above. 

In Table \ref{tab:q_inj_ky_1} we report the values of the injection
matrix computed both with our WFM method ($\mathbf{Q}_{inj}$) and
with the ``classical'' WFM by Komyakov \textit{et al.}\citep{khomyakov05}
($\mathbf{Q}_{inj}^{K}$) for the wave vector $k_{y}=1$. We considered
four frequencies: $\omega=0.50$, $\omega=1.00$, $\omega=1.55$ and
$\omega=2.20$. These are chosen so as to include both propagating
and evanescent modes. It is clear that, whenever one or both of the
modes are decaying, our method yields a vanishing matrix element corresponding
to the transmission from that mode. However, the classical method
always gives finite matrix elements. As a result, our method selects
from the outset only the transmission from a propagating mode. In
the classical method this is not the case, and to obtain the final
transmission amplitudes one has to zero out by hand the transmission
amplitude whenever ``incoming'' mode is decaying. 

We show in Fig. \ref{t_km_ky_0} the mode-resolved transmission amplitudes
for the case of $k_{y}=0$ as obtained by our method and as obtained
using the Fisher-Lee formula. It is evident that our results coincide
perfectly with those from the Fisher-Lee formula. In Fig. \ref{t_km_ky_1}
we show the same quantities computed for $k_{y}=1$. We now can see
that the results are very different than one another. Moreover, it
is striking to see that the transmission amplitudes computed using
the Fisher-Lee relation are in some cases greater than the maximum
allowed value of unity. The difference, as discussed above, appears
only when the wave vector matrices $\mathbf{C}_{L}^{a}$ and $\mathbf{C}_{R}$
are not orthogonal. This is, indeed, the case for the choice $k_{y}=1$.
We surmise that the failure of the Fisher-Lee relation to give proper
mode-resolved transmission amplitudes has eluded researchers because
they probably checked their results only in the particular situations
where the wave vector matrices are orthogonal; this is the case when
$k_{y}=0$ as shown above. And, more importantly, many such applications
were carried out in the one-dimensional case where such situation
is never encountered.

\section{Conclusion}

In conclusion, we have reformulated the conventional WFM method and
completed its equivalence with the AGF approach. Through a clever
choice of the evanescent eigenmodes entering the calculation, we need
only use four Bloch matrices, getting rid of the redundant four others
introduced in earlier literature, and which have no counterparts in
the AGF method. This choice also means that the injection matrix selects
from the outset the propagating modes as the sole that contribute
to the transmission. In order to arrive at the equivalence, we showed
how to obtain the group velocity matrices used in the WFM method from
the ``escape rates'' defined in the AGF formalism. This then allowed
us to proceed to a more physical spectral decomposition of the transmission
amplitude. The ``escape rates'' are related to the self-energies,
and these are computed differently in the WFM and in the AGF formalisms.
In the first, they are obtained from the Bloch matrices, and in the
second from the surface Green's functions of the leads. The findings
of this study intimate that when proper care is taken in the choice
of the phonon modes that enter the machinery of the WFM method, its
equivalence with the AGF technique is complete. As an illustration,
we applied our method to a simple model and, additionally, we made
a thorough comparison with earlier methods. Our results demonstrate
the equivalence of the two methods to a very high accuracy, limited
only by the machine precision. The comparison with the Fisher-Lee
relation sheds a light on its failure to properly decompose the transmission
into individual contributions of the phonon modes.

We have succeeded in completing the equivalence of the two most widely
used methods in ballistic transport at the nanoscale. Besides, we
have devised a procedure that provides a number of advantages when
compared to earlier approaches. The main advantages of our method
are as follows:
\begin{enumerate}
\item We used the inverse of the Bloch factor in the eigenvalue problem
of the left lead, conventionally chosen as the side from which the
incident phonons arrive. This choice precludes many costly matrix
inversions in the following steps of the procedure.
\item We defined four Bloch matrices, instead of eight as in Refs. \citep{khomyakov05,Ong2018},
which are enough to carry out all the subsequent calculations. 
\item Our injection matrix selects out the evanescent modes from the outset.
In prior methods \citep{khomyakov05}\textcolor{blue}{.} the latter
are eliminated ``by hand'' or by using a projector operator to project
out their unwanted contribution (Eq. 2.62 in \citep{khomyakov05}).
\item Each of the four Bloch matrices defined in this work has its equivalent
quantity in the AGF method, contrary to the eight matrices defined
in prior methods, where four of them have no corresponding quantities
in the AGF method.
\item The transmission function in our method is given in terms of the bulk
group velocities of the leads. These are guaranteed to be diagonal
and positive, and their square roots are clearly defined. However,
in prior methods, the transmission function is written in terms of
the ``escape rates''. These being defined as the difference between
a self-energy matrix and its conjugate, are not necessarily real as
is assumed in most publications \citep{Wang2008,Guevas1998,Guevas2017,Hafner2008}.
\item And of utmost importance, we derived a generalized Fisher-Lee formula
to properly decompose the transmission function into individual phonon
mode contributions, which is valid regardless of the orthogonality
of the eigenvectors. 
\end{enumerate}
These findings add to a growing body of literature on the theoretical
investigation of transport, thermal as well as electronic, through
interfaces at the nanoscale. The complete equivalence of the two most
widely used methods allows for a better understanding of the underlying
physics, especially in the AGF method, where, despite its obvious
mathematical superiority, lacks somewhat the transparency inherent
to the WFM method.

\ack{}{}

The authors would like to thank Professor J. K. Freericks for his
constructive criticism of this work and diligent proofreading of the
manuscript. This work was supported by the Algerian "Direction Générale de la Recherche Scientifique et de Développement Technologique" (DGRSDT).

\appendix

\section{Velocity matrices\label{app:velocity-matrices}}

Starting from the definition of the Bloch matrix $\mathbf{B}_{R}=\mathbf{C}_{R}\boldsymbol{\Lambda}_{R}\tilde{\mathbf{C}}_{R}$,
we have $\mathbf{B}_{R}\mathbf{C}_{R}=\mathbf{C}_{R}\boldsymbol{\Lambda}_{R}$
and $\mathbf{C}_{R}^{\dagger}\mathbf{B}_{R}^{\dagger}=\boldsymbol{\Lambda}_{R}^{\dagger}\mathbf{C}_{R}^{\dagger}$.
We make use of these equalities in the definition of the velocity
matrix (Eq. \ref{v_r_matrix}) and obtain
\begin{eqnarray}
\mathbf{V}_{R}= &  & \mathrm{i}\left[\mathbf{C}_{R}^{\dagger}\,\mathbf{H}_{RR}\,\mathbf{C}_{R}\,\mathbf{\boldsymbol{\Lambda}}_{R}-\mathbf{\boldsymbol{\Lambda}}_{R}^{\dagger}\,\mathbf{C}_{R}^{\dagger}\,\mathbf{H}_{RR}^{\dagger}\,\mathbf{C}_{R}\right]\nonumber \\
= &  & \mathrm{i}\left[\mathbf{C}_{R}^{\dagger}\,\mathbf{H}_{RR}\,\mathbf{B}_{R}\,\mathbf{C}_{R}-\mathbf{C}_{R}^{\dagger}\,\mathbf{B}_{R}^{\dagger}\,\mathbf{H}_{RR}^{\dagger}\,\mathbf{C}_{R}\right]\nonumber \\
= &  & \mathrm{i}\,\mathbf{C}_{R}^{\dagger}\left[\mathbf{H}_{RR}\,\mathbf{B}_{R}-\mathbf{B}_{R}^{\dagger}\,\mathbf{H}_{RR}^{\dagger}\right]\mathbf{C}_{R}\nonumber \\
= &  & \mathrm{i}\,\mathbf{C}_{R}^{\dagger}\left[\mathbf{\boldsymbol{\Sigma}}_{R}-\mathbf{\boldsymbol{\Sigma}}_{R}^{\dagger}\right]\mathbf{C}_{R}\nonumber \\
= &  & \mathrm{i}\,\mathbf{C}_{R}^{\dagger}\left[\mathbf{\boldsymbol{\Sigma}}_{R}-\mathbf{\boldsymbol{\Sigma}}_{R}^{a}\right]\mathbf{C}_{R}\nonumber \\
= &  & \mathbf{C}_{R}^{\dagger}\,\mathbf{\boldsymbol{\Gamma}}_{R}\,\mathbf{C}_{R},
\end{eqnarray}
where we used the expressions of the self-energy matrices in terms of the Bloch matrices as given in Eq. \ref{eq:WFM_R_self-energy}.

Starting again from the definition of the Bloch matrix $\mathbf{B}_{L}^{a}=\mathbf{C}_{L}^{a}\boldsymbol{\Lambda}_{L}^{a}\tilde{\mathbf{C}}_{L}^{a}$,
we have $\mathbf{B}_{L}^{a}\mathbf{C}_{L}^{a}=\mathbf{C}_{L}^{a}\boldsymbol{\Lambda}_{L}^{a}$
and $\mathbf{C}_{L}^{a\dagger}\mathbf{B}_{L}^{a\dagger}=\boldsymbol{\Lambda}_{L}^{a\dagger}\mathbf{C}_{L}^{a\dagger}$.
Similarly, we make use of these equalities in the definition of the
velocity matrix (Eq. \ref{vla_matrix}) and obtain

\begin{eqnarray}
\mathbf{V}_{L}^{a}= &  & -\mathrm{i}\left[\mathbf{C}_{L}^{a\dagger}\,\mathbf{H}_{LL}^{\dagger}\,\mathbf{C}_{L}^{a}\,\mathbf{\boldsymbol{\Lambda}}_{L}^{a}-\mathbf{\boldsymbol{\Lambda}}_{L}^{a\dagger}\,\mathbf{C}_{L}^{a\dagger}\,\mathbf{H}_{LL}\,\mathbf{C}_{L}^{a}\right]\nonumber \\
= &  & -\mathrm{i}\left[\mathbf{C}_{L}^{a\dagger}\,\mathbf{H}_{LL}^{\dagger}\,\mathbf{B}_{L}^{a}\,\mathbf{C}_{L}^{a}-\mathbf{C}_{L}^{a\dagger}\,\mathbf{B}_{L}^{a\dagger}\,\mathbf{H}_{LL}\,\mathbf{C}_{L}^{a}\right]\nonumber \\
= &  & -\mathrm{i}\,\mathbf{C}_{L}^{a\dagger}\left[\mathbf{H}_{LL}^{\dagger}\,\mathbf{B}_{L}^{a}-\mathbf{B}_{L}^{a\dagger}\,\mathbf{H}_{LL}\right]\mathbf{C}_{L}^{a}\nonumber \\
= &  & -\mathrm{i}\,\mathbf{C}_{L}^{a\dagger}\left[\mathbf{\boldsymbol{\Sigma}}_{L}^{a}-\mathbf{\boldsymbol{\Sigma}}_{L}^{a\dagger}\right]\mathbf{C}_{L}^{a}\nonumber \\
= &  & -\mathrm{i}\,\mathbf{C}_{L}^{a\dagger}\left[\mathbf{\boldsymbol{\Sigma}}_{L}^{a}-\mathbf{\boldsymbol{\Sigma}}_{L}\right]\mathbf{C}_{L}^{a}\nonumber \\
= &  & \mathbf{C}_{L}^{a\dagger}\mathbf{\boldsymbol{\Gamma}}_{L}\,\mathbf{C}_{L}^{a},\label{vla_app}
\end{eqnarray}
where we used the expressions of the self-energy matrices in terms
of the Bloch matrices as given in Eq. \ref{appb-2-3}.

\section{Injection matrix\label{app:injection-matrix}}

From Eq. \ref{Q_0_expression}, we have $\mathbf{Q}_{\mathrm{inj}}=\mathbf{H}_{LL}^{\dagger}(\mathbf{B}_{L}^{a}-\mathbf{B}_{L})\mathbf{C}_{L}^{a}$ and from Eqs. \ref{eq:WFM_L_self-energy} and \ref{appb-2-3}, we have
\begin{equation}
\mathbf{Q}_{\mathrm{inj}}=(\boldsymbol{\Sigma}_{L}^{a}-\boldsymbol{\Sigma}_{L})\mathbf{C}_{L}^{a}=i\boldsymbol{\Gamma}_{L}\mathbf{C}_{L}^{a}.\label{eq:Qinj_appendix}
\end{equation}
Then, from Eq. \ref{gamma_l}, the injection matrix reads
\[
\mathbf{Q}_{\mathrm{inj}}=i\left(\mathbf{C}_{L}^{a\dagger}\right)^{-1}\mathbf{V}_{L}^{a}.
\]

\section{Eigenvector matrices from the Green's functions\label{app:eigenvector-matrices}}

From Eqs. \ref{eq:B_La_WFM} and \ref{eq:B_R_WFM}, we multiply by
$\mathbf{C}_{L}^{a}$ and $\mathbf{C}_{R}$ on the right and obtain
\[
\mathbf{B}_{L}^{a}\mathbf{C}_{L}^{a}=\mathbf{C}_{L}^{a}\boldsymbol{\Lambda}_{L}^{a}
\]
and
\[
\mathbf{B}_{R}\mathbf{C}_{R}=\mathbf{C}_{R}\boldsymbol{\Lambda}_{R}.
\]
Also from the expressions of $\mathbf{B}_{L}^{a}$ and $\mathbf{B}_{R}$in
terms of the surface Green's functions(Eqs. \ref{appa-2-2} and \ref{appb-2-2-2}),
we obtain the following right eigenvalue equation for the eigenvector
matrices $\mathbf{C}_{L}^{a}$ and $\mathbf{C}_{R}$:
\[
\begin{array}{ccc}
\mathbf{g}_{L}^{a}\mathbf{H}_{LL}\mathbf{C}_{L}^{a} & = & \mathbf{C}_{L}^{a}\boldsymbol{\Lambda}_{L}^{a},\\
\mathbf{g}_{R}\mathbf{H}_{RR}^{\dagger}\mathbf{C}_{R} & = & \mathbf{C}_{R}\boldsymbol{\Lambda}_{R}.
\end{array}
\]
These are similar to the expressions found by Ong \citep{Ong2018}\textcolor{blue}{.}

\end{document}